\documentclass[journal, one column,single spacing ,12pt]{IEEEtran}
\IEEEoverridecommandlockouts
\usepackage{rotating,setspace,latexsym,amsmath,amssymb,bm}
\usepackage{cite}
\pdfoutput=1
\usepackage{algpseudocode}
\usepackage{algorithm}
\usepackage{algorithmicx}
\usepackage{xfrac}
\usepackage{amsthm}
\usepackage{subfig}
\usepackage{graphicx}
\usepackage{booktabs} 
\usepackage{caption}
\usepackage{amsmath}
\newtheorem{Thrm}{Theorem}
\newtheorem{exmpl}{Example}

\newtheorem{Lem}{Lemma}

\newtheorem{Def}{Definition}

\title{On the Latency and Energy Efficiency of Erasure-Coded Cloud Storage Systems}
\author{{Akshay Kumar, Ravi Tandon, T. Charles Clancy}
\thanks{A. Kumar and T. C. Clancy are with the Hume Center for National Security and Technology and Dept. of Electrical and Computer Engineering, Virginia Tech, Blacksburg, VA USA. Email: \{akshay2, tcc\}@vt.edu. R. Tandon is with the Discovery Analytics Center, Dept. of Computer Science,	Virginia Tech, Blacksburg, VA USA. Email: tandonr@vt.edu}
\thanks{Parts of this work were presented at the Globecom 2014 conference \cite{Kumar14}.
}}

\begin{document}
\maketitle 
\vspace{-30pt} 
\begin{abstract}
\vspace{-7pt}
The increase in data storage and power consumption at data-centers has made it imperative to design energy efficient Distributed Storage Systems (DSS). The energy efficiency of DSS is strongly influenced not only by the volume of data, frequency of data access and redundancy in data storage, but also by the heterogeneity exhibited by the DSS in these dimensions. To this end, we propose and analyze the energy efficiency of a heterogeneous distributed storage system in which $n$ storage servers (disks) store the data of $R$ distinct classes. Data of class $i$ is encoded using a $(n,k_{i})$ erasure code and the (random) data retrieval requests can also vary across classes. We show that the energy efficiency of such systems is closely related to the average latency and hence motivates us to study the energy efficiency via the lens of average latency. Through this connection, we show that erasure coding serves the dual purpose of reducing latency and increasing energy efficiency. We present a queuing theoretic analysis of the proposed model and establish upper and lower bounds on the average latency for each data class under various scheduling policies. Through extensive simulations, we present qualitative insights which reveal the impact of coding rate, number of servers, service distribution and number of redundant requests on the average latency and energy efficiency of the DSS. 
\end{abstract}
\vspace{-10pt}
\begin{IEEEkeywords}
\vspace{-5pt}
Erasure Codes, Distributed Storage, Fork-Join Queues, Latency, Energy Efficiency, Multi-class queuing system.
\end{IEEEkeywords}
\section{Introduction}
Cloud based storage systems are emerging to gain significant prominence due to their highly virtualized infrastructure that presents cost-effective and simple to use elastic network resources. The backbone infrastructure of the cloud is comprised of distributed storage systems (DSS), in which the data is stored and accessed from commodity storage disks. Coding of data across distributed disks provides fault tolerance by providing reliability against unexpected disk failures. There has been a recent paradigm shift from classical replication based codes to erasure codes because they provide higher fault tolerance at the same storage cost \cite{Weatherspoon02}. As a result, a number of commercial DSS such as Google Colossus, Windows Azure etc. are transitioning to the use of erasure codes \cite{Colossus, Facebook, Azure}. Besides providing fault tolerance and minimizing storage cost, another important aspect which deserves equal, if not more attention is the \textit{energy efficiency} of DSS. 

Over the last decade, the dramatic usage of data has lead to an enormous increase in the volume of stored (archival) data and the frequency of data access to a DSS \cite{dataIncrease1}. This translates to more and more servers being added to the data-center operating at higher server utilization levels. As a result, the energy consumption of the data-centers is increasing steeply and adds up to its operational cost. According to \cite{Sverdlik}, energy consumed by the data centers globally has increased by 19\% in 2012 and storage systems in a large data-center consume up to 40\% of the total energy \cite{HarnikNaor09}. Hence, there is a need to devise energy efficient data storage schemes. The existing techniques for energy-efficient data storage are based on variants of schemes that involve powering off storage devices \cite{MAID, SRCMap, JoKwon10}. 

Energy efficiency is a system wide property and while some metrics focus on the energy efficiency of hardware or software components \cite{chen11}, others are based on the usage of physical resources (such as CPU, memory, storage etc.) by the running applications or servers. For the scope of this work, we focus on the \emph{data transfer throughput metric} \cite{Schulz}, which measures energy efficiency as the amount of data processed in the DSS per unit amount of energy expended across all distributed servers. Therefore, the energy efficiency of DSS is strongly influenced by the volume of data transferred (per request), frequency of data storage/access requests, service rate of each server and the degree of redundancy in data storage.

The energy efficiency of a DSS is also closely related to its read/write latency\footnote{Here, latency refers to the time taken to process a data request, measured relative to the time at which it enters the DSS. For the scope of this work, we consider latency to be the sum of queuing delay and service time, and assume the other delays to be relatively negligible.}. The data stored and accessed from the cloud is broadly classified into two categories \cite{hotCold1, hotCold2}:
\begin{itemize}
\item \emph{Hot-data}: this could refer to data which is frequently accessed (i.e., a higher job request rate). Furthermore, it is desirable to provide higher redundancy/fault tolerance when storing such data.
\item \emph{Cold-data}: this could refer to data which is infrequently accessed or archival data. Such data does not necessarily mandate to be coded and stored with higher fault tolerance, as it is seldom accessed by the users.
\end{itemize}
When the data is infrequently accessed as in case of \emph{Cold-data}, the average latency is reduced and it also improves the energy efficiency of DSS \cite{richardPhD13}. Another case in point is that increasing redundancy as in \emph{Hot-data} improves fault-tolerance but generally results in increased latency \cite{gauri13}. The energy efficiency in this case decreases due to increase in power consumption as more servers are involved in processing the same data request. Thus the latency of DSS is closely tied with its energy efficiency. Therefore in this work, we study the energy efficiency of a DSS through the lens of average latency of DSS.

As mentioned earlier, the erasure coded DSS, due to their several merits over replication based codes, have gained significant prominence in recent times. Therefore, in this work we study the relationship between latency and energy efficiency for such systems. In a erasure coded DSS, the data of each user is stored across $n$ disks (or servers) using a $(n, k)$ optimal Maximum-Distance-Separable (MDS) code. By the property of MDS codes, accessing the data stored at {\textit{any}} $k$ out of $n$ servers suffices to recover the entire data of a user (also referred to as successful completion of the {\textit{job request}}\footnote{We restrict our attention to read requests because in most of the practical DSS, such as HDFS \cite{hdfs}, Windows Azure \cite{Azure} etc. the user's data is written only once to the storage nodes but it can be retrieved multiple times by the user.} of that user). The processing of job requests in DSS is typically analyzed using Fork-Join (F-J) queues \cite{conway63, dijkstra68}. A $(n,k)$ F-J queue consists of $n$ independently operating queues corresponding to each of the $n$ servers. Every job arriving in the system is split $n$ ways and enters the queues of all $n$ servers simultaneously. A queuing theoretic latency analysis of the $(n,k)$ F-J system has been done in \cite{gauri13} (also see \cite{longbo12, shahLee12, Shah13}). The key findings of these papers is that using erasure coding and sending redundant requests (requests to more than $k$ servers for a $(n,k)$ F-J system) can significantly reduce the latency of a DSS. 
 
However, most of the aforementioned literature considers a homogenous storage architecture and there is no distinction (from system's perspective) between any two job requests entering the system. However, that is hardly the case with real DSS, wherein as mentioned earlier (see \emph{Hot-data} vs. \emph{Cold-data}), the job requests can be classified into one of the several classes based on the job arrival rate or fault-tolerance/storage requirements. For instance, the leading cloud storage providers such as Amazon S3, Windows Azure etc. allow their customers to choose from multiple storage classes that differ in redundancy/fault-tolerance, data availability and access latency \cite{AmazonPricing,googlePricing}. They have a low redundancy storage class designed for infrequently/archival data. Therefore, motivated by this fact, we consider a $(n, k_1, k_2, ..., k_R)$ multi-tenant DSS for $R$ distinct data classes, a generalization of the homogenous $(n,k)$ DSS in \cite{gauri13}. Data of class $i$ $(\forall i \in \{1, 2, ..., R\})$ is stored across $n$ servers using a $(n,k_i)$ erasure (MDS) code. The arrivals\footnote{Job arrivals refers to the time instants at which job requests enters the queues of the servers in the DSS.} of job requests of class $i$ are assumed to follow a Poisson distribution with rate $\lambda_i$. 
 
The key contributions of this paper are:
\begin{itemize}
\item A multi-tenant DSS is proposed and analyzed through the Fork Join framework to account for the heterogeneity in job arrival rates and fault-tolerance requirements of different data classes.
\item A \emph{data throughput} based energy efficiency metric is defined for the heterogeneous DSS operating under any given scheduling policy. For the special case of single server and data class, we showed that the average latency and energy efficiency of DSS are closely related to each other. Therefore, using a queuing-theoretic approach, we provided lower and upper bounds on the average latency for jobs of class $i$ ($\forall i \in \{1, 2, ..., R\}$) in the proposed F-J framework under various scheduling policies such as First-Come-First-Serve (FCFS), preemptive and non-preemptive priority scheduling policies.
\item We studied the impact of varying the code-rate on the latency, energy efficiency and network bandwidth consumed by DSS. Increasing code-rate reduces latency and increases energy efficiency. However, this comes at the cost of increased storage space and (write) bandwidth. We also obtained interesting insights from investigating the impact of varying the number of servers, heavy-tail arrival/service distributions in the DSS.
\item Lastly, we studied the impact of varying the number of redundant requests (sending requests to more than $k$ servers for $(n,k)$ MDS code) to the DSS. We observed that sending redundant requests reduces latency and increases energy efficiency. Thus, full redundancy results in minimum latency and maximum energy efficiency for each data-class.
\end{itemize}

\section{Related Work}
\label{litReview}
A number of good MDS codes such as LINUX RAID-6 and array codes (EVENODD codes, X-code, RDP codes) have been developed to encode the data stored on cloud (see \cite{Plank} and references therein). These codes have very low encoding/decoding complexity as they avoid Galois Field arithmetic (unlike the classical Reed-Solomon MDS codes) and involve only XOR operations. However, they are usually applicable upto two or three disk failures. Also, in the event of disk failure(s), Array codes and recently introduced Regenerating codes reduce disk and network I/O respectively. Recently, non-MDS codes such as Tornado, Raptor and LRC codes \cite{LTcode, raptorCodes} have been developed for erasure coded storage. Although the fault-tolerance is not as good as MDS codes, they achieve higher performance due to lower repair bandwidth and I/O costs. 

The latency analysis of (MDS) erasure coded $(n,k)$ homogenous DSS has been well investigated in \cite{gauri13, longbo12, shahLee12} which provide queuing theoretic bounds on average latency. A related line of work \cite{Chen14, Shah13} independently showed that sending requests to multiple servers always reduces the (read) latency. Then Liang et. al. \cite{liangKozat} extended the latency analysis to a $(n,k,L)$ DSS, in which $n$ of a total $L$ number of independent servers are used to store the $(n,k)$ MDS code. It assumed a ``constant+exponential'' model for the service time of jobs. The authors in \cite{liang14, liangKozat14} developed load-adaptive algorithms that dynamically vary job size, coding rate and number of parallel connections to improve the delay-throughput tradeoff of key-value storage systems. These solutions were extended for  heterogeneous services with mixture of job sizes and coding rate. Recently, Xiang et. al. \cite{Vaneet14} provided a tight upper bound on average latency, assuming arbitrary erasure code, multiple file types and a general service time distribution. This was then used to solve a joint latency and storage cost optimization problem by optimizing over the choice of erasure code, placement of encoded chunks and the choice of scheduling policy.  

Data-centers while configured for peak-service demand, end up being highly underutilized. Furthermore, the hardware components in storage systems are not power proportional, with idle mode consuming roughly 60\% of that of a busy power \cite{Barroso07}. This has resulted in significant research in designing/implementing power efficient schemes. Most of the current literature focuses on power management via performance scaling (such as DVFS \cite{Snowdon05, Andrew10}) or low-power states \cite{Meisner09}. Recently, Liu et. al. \cite{Liu13} investigated the effect of varying various system operations such as processing speed, system on/off decisions etc. on the power-delay performance from a queuing theoretic perspective. This work was extended in \cite{Liu14}, wherein a joint speed scaling and sleep state management approach was proposed, that determines the best low-power state and frequency setting by examining the power consumption and average response time for each pair.

However, the work in \cite{Liu13, Liu14} does not present the power-delay performance analysis in a erasure coded DSS. Also it focuses on power consumption rather than the more relevant, energy efficiency of DSS. Therefore, in this work, we study the relationship between energy efficiency and average latency in a (MDS) erasure coded heterogeneous DSS for different scheduling policies.

\section{System Model}
\label{sysModel}
\begin{figure}[!t]%
\centering
\includegraphics[scale=0.35]{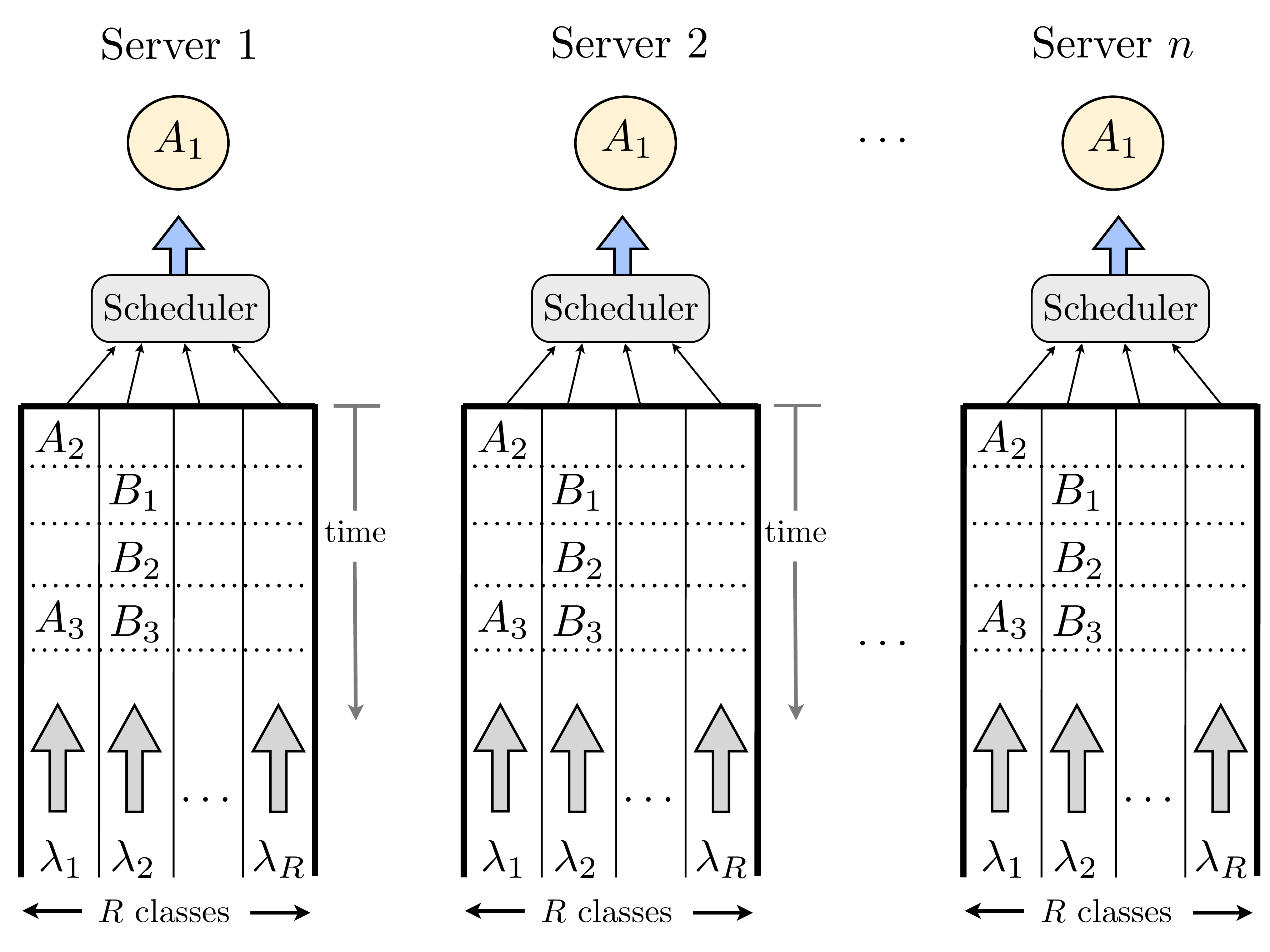}%
\vspace{0pt}
\caption{System Model.}%
\label{system_Model}%
\end{figure}

A heterogeneous multi-tenant $(n, k_1, k_2, ..., k_R)$ DSS (shown in Fig.~\ref{system_Model}) consists of $n$ servers that store the data of $R$ distinct classes. The $R$ classes differ from each other in the fault-tolerance, storage requirements and frequency of access of the stored data. The data of class $i$ (which is assumed to be of size $l_i$) is partitioned into $k_i$ equal size fragments and then stored across $n$ servers using a $(n,k_i)$ Maximum-Distance-Separable (MDS) code. Thus each server stores, $1/k_i$ fraction of original data. The arrival process for request of class $i$ is assumed to be Poisson with rate $\lambda_i$. The service time at each server is assumed to follow an exponential distribution with service rate $\mu$ (per unit file size) \cite{liangKozat14}. The effective service rate at any server for jobs of class $i$ is $\mu_i = \frac{k_i \mu}{l_i}$ since each server stores $1/k_i$ fraction of data.
\vspace{-5pt}
\begin{exmpl}
\label{example}
We now present a representative example to illustrate the system model. Consider a $(n,k_1,k_2) = (3,2,1)$ two-class DSS. Data for the two classes $A$ and $B$ are encoded across $n=3$ servers using $(3,2)$ and $(3,1)$ MDS codes respectively as shown in Fig.~\ref{mdsCode}. Let $A_1$ and $B_1$ denote two files of class $A$ and $B$ respectively that need to be coded and stored across the servers. Then for the $(3,2)$ MDS code, $A_1$ is split into two sub-files, $A_{11}$ and $A_{12}$, of equal size and are stored on any two servers (servers 1 and 2 in Fig.~\ref{mdsCode}). Then the remaining server (i.e.\ server 3) stores $A_{11} \oplus A_{12}$. Thus each server stores half the size of original file and the entire file can be recovered from any two servers. The $(3,1)$ MDS code for file $B_1$, is a simple replication code in which each server stores the copy of entire file of class $B$ and thus can be recovered by accessing the data from any one server. 
\begin{figure}[!t]%
\centering
\includegraphics[scale=0.35]{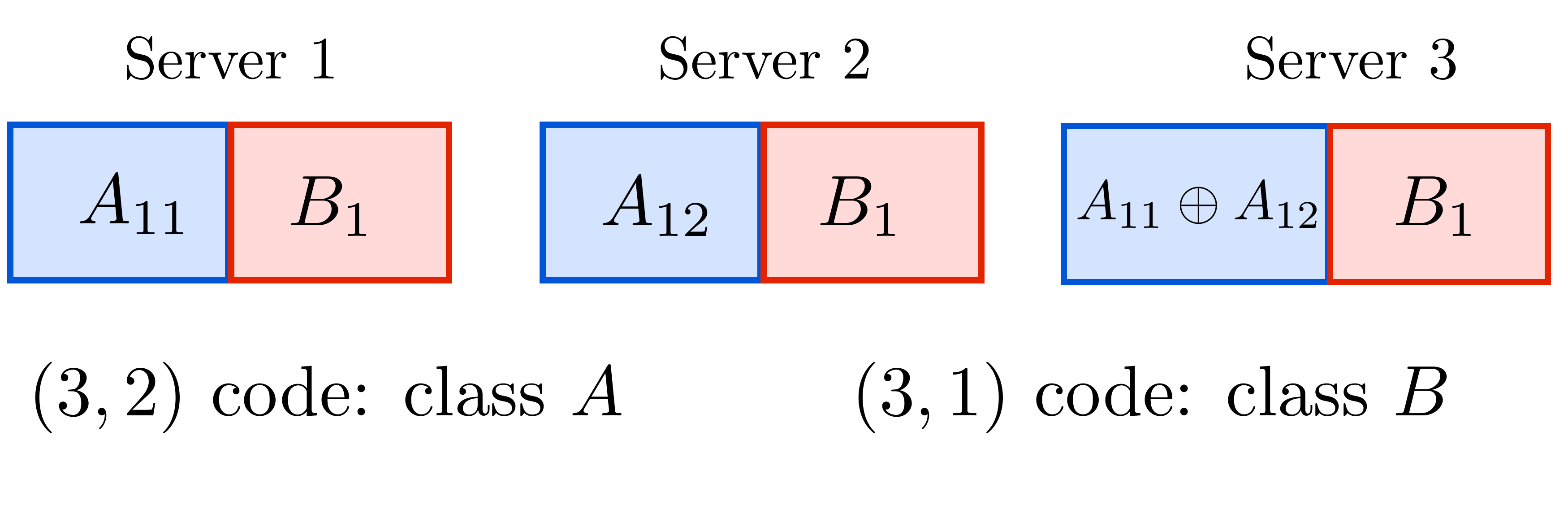}%
\vspace{-10pt}
\caption{MDS codes for data storage in a two-class Fork-Join system.}%
\label{mdsCode}%
\vspace{10pt}
\end{figure}
\end{exmpl}
The evolution of system state in this example, depends on the local scheduling policy at each server. Although there exists various scheduling policies, in this work we consider First-Come-First-Serve (FCFS), preemptive and non-preemptive priority queuing policies at each server. In FCFS scheduling, all data classes are equal priority. At each server, the job that enters first in the buffer is served first. In a priority queuing policy, the data classes are assigned different priority levels. A job of a particular class will be served only when there are no outstanding jobs of classes with higher priority level. A priority queuing policy is further classified as preemptive or non-preemptive based on whether or not the job in server can be preempted by a job of higher priority level.   

\begin{figure}[!t]%
\centering
\includegraphics[scale=0.5]{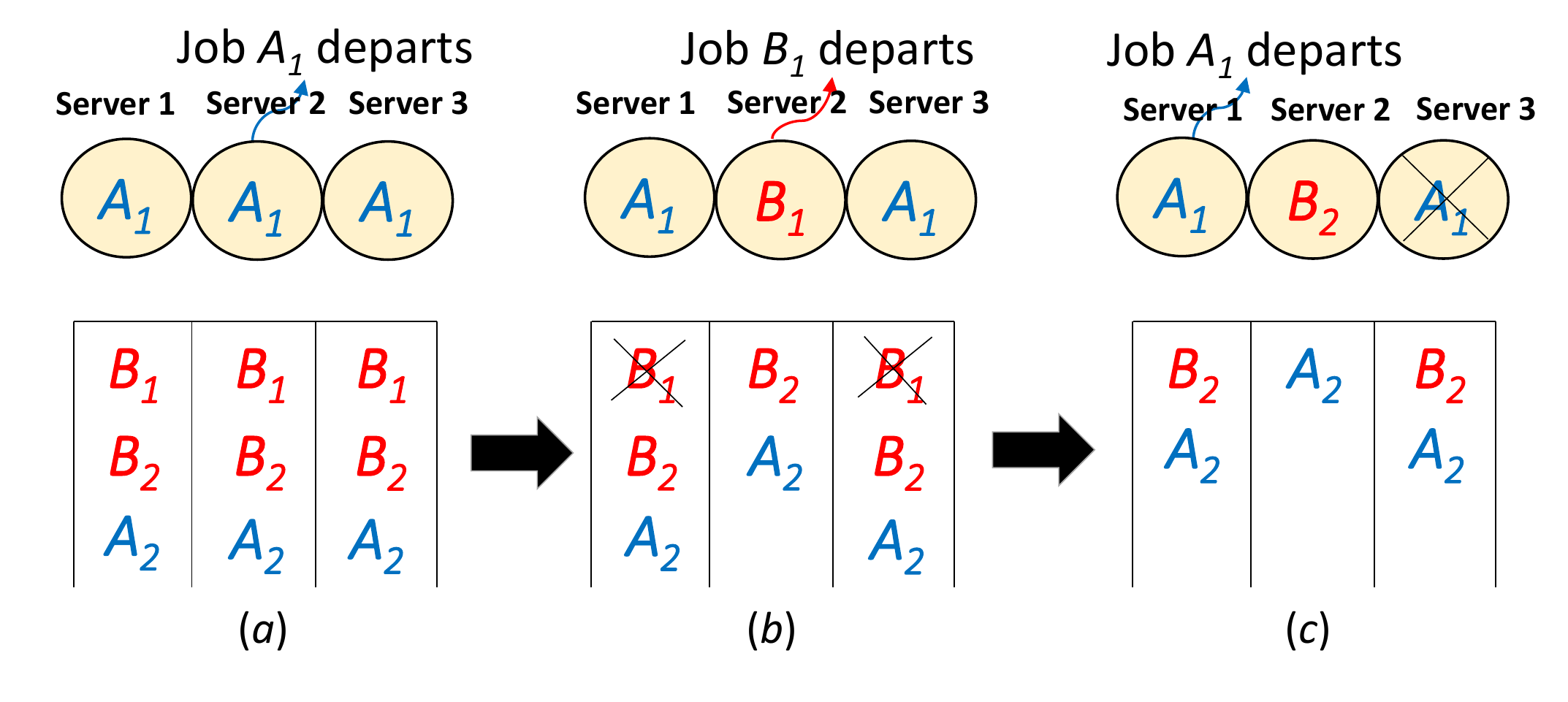}%
\caption{System state evolution: two-class FJ system with FCFS.}%
\vspace{4pt}
\label{fcfs}%
\end{figure}

Fig.~\ref{fcfs}(a)-(c) illustrates the evolution of system state under the FCFS policy.   After server 2 finished job $A_1$ in Fig.~3(a), $B_1$ enters server 2 and is finished in the next state (Fig.~3(b)) while other servers still process $A_1$. Since $k_B=1$, the remaining two copies of $B_1$ immediately exit the system. Finally in Fig. 3(c) server 1 finishes $A_1$ and since $k_A = 2$, $A_1$ exits at server 3. 

\subsection{Latency and Energy Efficiency}
\begin{Def}\label{latencyDefn}
For a $\{\lambda_i, l_i, \mu, (n, k_1, k_2, ..., k_R)\}$ DSS, the average latency of class $i$ under some scheduling policy $\mathcal{P}$ is defined as
\begin{equation}
T^i_{\mathcal{P}} = T^i_{s,\mathcal{P}} + T^i_{q,\mathcal{P}},
\label{DefLatency}
\end{equation}
where $T^i_{s,\mathcal{P}}$ and $T^i_{q,\mathcal{P}}$ are the average service time and waiting time (in queue) for a job of class $i$ respectively.
\end{Def}

\begin{figure}%
\includegraphics[scale=0.50]{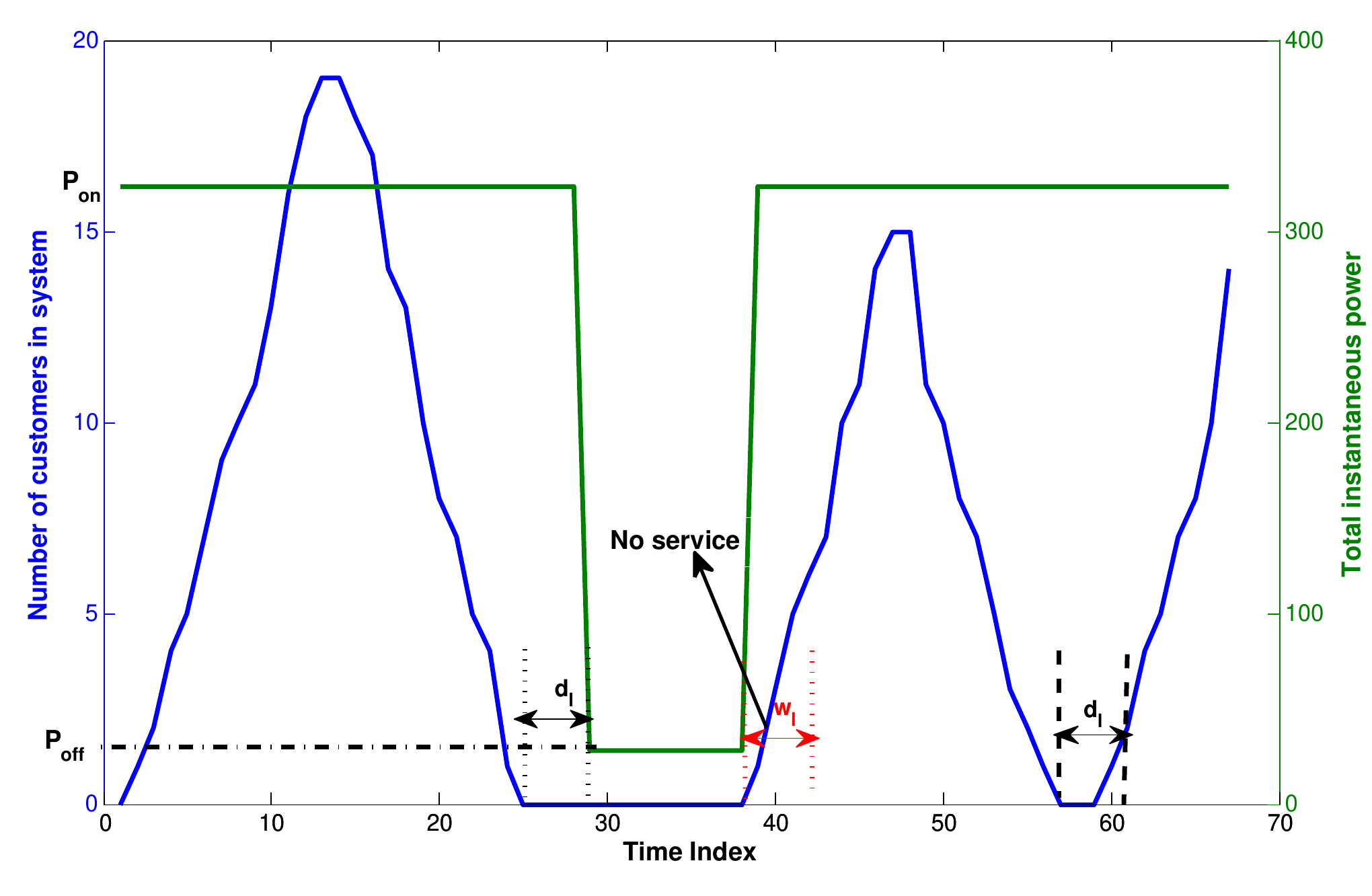}%
\centering
\caption{\footnotesize{Variation of total power consumption of DSS across multiple busy periods and idle periods. The switch to idle state happens only for idle periods with duration greater than $d_l$ but it then results in a wake-up latency of $w_l$.}}%
\label{systemModel}%
\end{figure}

For a $\{\lambda_i, l_i, \mu, (n, k_1, k_2, ..., k_R)\}$ DSS operating under scheduling policy $\mathcal{P}$, a relevant metric for measuring the energy efficiency, $\mathcal{E}_{\mathcal{P}}$, of the DSS is the data transfer throughput metric \cite{Schulz}. It is defined as the limiting ratio of the amount of data processed, $D_{\mathcal{P}}(t)$, by the DSS to the energy consumed, $E_{\mathcal{P}}(t)$, by the DSS in a infinitely large time duration $t$. It has units of bits/Joule. Now $D_{\mathcal{P}}(t)$ is simply,
\begin{equation}
D_{\mathcal{P}}(t) = \sum_{i=1}^R l_i N_i(t),
\label{dataProc}
\end{equation}
where $N_i(t)$ is the number of jobs of class $i$ processed by DSS in a time interval $t$. In order to determine $E_{\mathcal{P}}(t)$, we model the power consumption of the DSS as follows:
\begin{itemize}
\item To reduce power consumption, the servers are equipped with the dynamic voltage/frequency scaling (DVFS) mechanism and low-power states \cite{Liu14}. The DVFS mechanism reduces operating voltage and CPU processing speed (or frequency) in step to reduce utilization and hence increase power savings. 
\item The power consumed by a server in any state is the sum of power consumed by the CPU and the platform which comprises of chipset, RAM, HDD, Fan etc. The power consumed by the CPU and platform in a given state is assumed to be same across all the $n$ servers.
\item The power consumed by a server (CPU and platform) while being in active and low-power state is denoted by $P_{\text{on}}$ and $P_{\text{off}}$ respectively. A server is in active mode during the busy periods (i.e., there are outstanding jobs waiting for service). In general, at the end of a busy period, a server remains active for a while and then enters a sequence of low-power states staying in each for a predetermined amount of time. For ease of analysis, we lump them into a single low-power state with constant CPU power, $C_l$ and constant platform power, $P_l$. After the busy period is over, the server remains in active mode for $d_l$ and then enters the low-power state\footnote{As a consequence, if the duration of idle period (time between end of a busy period and start of the next one) is smaller than $d_l$, then the server always remains active).}. When the busy period restarts, the server incurs a wake-up latency $w_l$ in which it consumes active mode power, but is not capable of processing any job requests. Fig.~\ref{systemModel} explains this using an example.
\item The CPU power during active mode, $C_a$ is proportional to $V^2 f$, where $V$ is the supply voltage and $f$ is the CPU operating frequency\footnote{Due to this, the effective service rate for class $i$ becomes $\mu_i= f k_i \mu/l_i$.} ($f\in [0, 1]$) and are set by the DVFS mechanism. Further, we assume that $V$ is proportional to $f$ \cite{Liu14}. So $C_a=C_0 f^3$, for some maximum power $C_0$. The power consumed by the platform during active mode, $P_a$, is constant.
\item $t_{\text{busy}}^{i, j, k}$ denotes the duration of time for which the $k^{\text{th}}$ server is \emph{busy} serving $j^{\text{th}}$ job of $i^{\text{th}}$ class. 
\item $t_{\text{idle}}^{i, j, k}$ denotes the duration of idle period after the $k^{\text{th}}$ server finished the $j^{\text{th}}$ job of $i^{\text{th}}$ class.
\end{itemize}
Using the above notations, the active mode power per server is $P_{\text{on}} = C_a+P_a = C_0 f^3+P_a$. Similarly, $P_{\text{off}} = C_l+P_l$. Consider any time duration $t$ of interest during the operation of DSS. During this period, the total time for which the DSS is in active mode, $t_a$, is sum total (across all servers) of all busy periods plus the active mode time before entering low-power state. Mathematically, we have,
\begin{equation}
t_a = \sum_{i=1}^R \sum_{j=1}^{N_i(t)} \sum_{k=1}^{n} t_{\text{busy}}^{i,j,k} + \text{max}~(0,t_{\text{idle}}^{i, j, k}-d_l)
\label{activeTime}
\end{equation}
The total time for which DSS is in low-power state, $t_l$ is,
\begin{equation}
t_l = nt-t_a.
\label{idleTime}
\end{equation}
We have now the following definition of energy efficiency of a DSS. 
\begin{Def}\label{energyEffDefn}
For a $\{\lambda_i, l_i, \mu, (n, k_1, k_2, ..., k_R)\}$ DSS, the energy efficiency of the DSS under some scheduling policy $\mathcal{P}$ is defined as,
\begin{align}
\mathcal{E}_{\mathcal{P}} &= \lim_{t\rightarrow \infty} \frac{D_{\mathcal{P}}(t)}{E_{\mathcal{P}}(t)}, \label{DefenergyEff1}\\
 &= \lim_{t\rightarrow \infty} \frac{\sum_{i=1}^R l_i N_i(t)}{P_{\text{on}} t_a + P_{\text{off}} t_l}, \label{DefenergyEff2}
 %&= \lim_{t\rightarrow \infty} \frac{\sum_{i=1}^R l_i N_i(t)}{(P_{\text{on}}-P_{\text{off}})\sum_{i=1}^R \sum_{j=1}^{N_i(t)}\sum_{k=1}^{n} t_{\text{busy}}^{i,j,k}+\text{max}~(0,t_{\text{idle}}^{i, j, k}-d_l) + P_{\text{off}} nt},\\
\end{align}
\end{Def}
where \eqref{DefenergyEff2} follows from \eqref{DefenergyEff1} using \eqref{dataProc}. The expressions for $t_a$ and $t_l$ are given in \eqref{activeTime} and \eqref{idleTime} respectively.
%Fig.~\ref{systemModel} illustrates the energy efficiency metric through an example. It shows the dynamics of power consumed at a server processing job requests of two distinct classes (shown in blue and red) for low and high values of job arrival and service rates. During a given time interval $t$, the number of jobs processed of a class $i$, $N_i(t)$ increases on increasing both the arrival and service rates. However, this also changes the busy time ($t_{\text{on}}$) of servers.
%\begin{figure}%
%\includegraphics[width=\columnwidth]{energyEff.pdf}%
%\caption{Example illustrating the energy efficiency of a DSS with two classes (shown in red and blue). (a) low $\lambda_1, \lambda_2, \mu$ (b) high $\lambda_1, \lambda_2, \mu$.}%
%\label{energyEff}%
%\end{figure}

Next in order to highlight the relationship between the average latency and energy efficiency of a DSS, we consider the special case of a M/M/1 system and a single data-class. For tractability of analysis, here we assume that $d_l$, $w_l$ and $P_{\text{off}}$ are all $0$. Then from the Definition~\ref{latencyDefn} for average latency\footnote{In this special case, the scheduling policy $\mathcal{P}$ and class index $i$ are not relevant and hence dropped from \ref{DefLatency}.}, we have,
\begin{align}
T &= T_s + T_q, \label{latencyMM1}\\
 &= \frac{1}{\mu^{'}} + \frac{\lambda}{\mu^{'}(\mu^{'}-\lambda)}=\frac{1}{\mu^{'}-\lambda}, \label{latencyTerms}
\end{align}
where \eqref{latencyTerms} follows from \eqref{latencyMM1} by noting that for a M/M/1 system, the mean service time is $T_s = \frac{1}{\mu^{'}}$ and mean waiting time is $T_q=\frac{\lambda}{\mu^{'}(\mu^{'}-\lambda)}$. Here, $\mu^{'} = \frac{\mu f}{l}$ is the effective service rate. Here,
The energy efficiency is computed using \eqref{DefenergyEff2} as
\begin{align}
\mathcal{E} &= \lim_{t\rightarrow \infty} \frac{l N(t)}{P_{\text{on}} t_{\text{a}} + P_{\text{off}} t_{\text{l}}}, \label{energyEff1}\\
&= \lim_{t\rightarrow \infty} \frac{l N(t)}{P_{\text{on}} \sum_{i=1}^{N(t)} T_{s,i}}, \label{energyEff2}\\
&= \frac{l}{P_{\text{on}} \lim_{t\rightarrow \infty} \frac{\sum_{i=1}^{N(t)} T_{s,i}}{N(t)}}, \label{energyEff3}\\
&= \frac{l}{P_{\text{on}} T_{s}} \label{energyEff4},
\end{align}
where \eqref{energyEff2} follows from \eqref{energyEff1} by noting that $t_{\text{on}}$ is sum of service time of each of $N(t)$ jobs (denoted by $T_{s,i}$ for the $i^{\text{th}}$ job) and by neglecting the power consumed when server is idle i.e., $P_{\text{off}}=0$. Then \eqref{energyEff4} follows from \eqref{energyEff3} from the definition of average service time. Thus the energy efficiency is inversely related to the average service time of jobs. It is difficult to find a closed form expression for the energy efficiency of a heterogeneous DSS but the general trend of inverse proportionality between latency and energy efficiency continues to hold true as verified through extensive simulations in Section~\ref{result}. The average latency is also directly related to the average service time\footnote{Queuing delay depends on job arrival rate and service time. So the latency which is sum of queuing delay and service time directly depends on service time.}. Therefore, we conclude that energy efficiency and average latency of a DSS are closely related to each other. Henceforth, we focus on the latency analysis of a heterogeneous DSS.

\section{Preliminaries}
\label{prelim}
In this section, we first present the analysis of average service latency in a multi-class single server system with FCFS scheduling policy. For the corresponding results in a priority (preemptive/non-preemptive) queuing system, we refer the reader to \cite{gallager}. To improve the tractability of the latency analysis, the analytical results in this work ignore the impact of wakeup latency $w_l$ similar to other works in literature \cite{gauri13, longbo12, shahLee12, Chen14, Shah13}. We then briefly review the existing results for upper and lower bounds on the average latency for a $(n,k)$ homogenous Fork-Join system \cite{gauri13}. 

\subsection{Average Latency in Multi-class Single Server System with FCFS Scheduling} 
\label{respTimeSingleServer}
Consider the system model described in Fig.~\ref{system_Model} with $n=1$ server and FCFS scheduling policy. The FCFS system can be modeled as a M/G/1 queuing system with net arrival rate, $\lambda = \sum\limits_{r=1}^R{\lambda_r}$, and a general service distribution, $S$. The average latency of jobs of class $i$ is the sum of their average waiting time (in queue) and the average service time. Let $S_i$ be a random variable representing the service time for a job of class $i$ in the FCFS system. Then the average service time of jobs of class $i$ is simply the expectation, $\textnormal{E}[S_i]$. In the FCFS system, the waiting time, $W_{\textnormal{FCFS}}$, for jobs of all the classes is same and is given by the Pollaczek-Khinchine (P-K) formula \cite{tijms03} (for M/G/1 system) as 
\begin{align}
W_{\textnormal{FCFS}} = \frac{\lambda (\textnormal{E}[S^2])}{2(1-\lambda \textnormal{E}[S])}, \label{waitFCFS}
\end{align}
Therefore, the average latency for jobs of class $i$ is, 
\begin{align}
T_{\textnormal{FCFS}}^i = \textnormal{E}[S_i]+ \frac{\lambda \textnormal{E}[S^2]}{2(1-\lambda \textnormal{E}[S])}= \textnormal{E}[S_i]+ \frac{\lambda (\textnormal{V}[S]+{\textnormal{E}[S]}^2)}{2(1-\lambda \textnormal{E}[S])}, \label{eq:timeFCFS} 
\end{align}
where $\textnormal{V}[.]$ denotes the variance of the random variable.
Now the fraction of jobs of class $i$, $p_i$ is
\begin{align}
p_i = \frac{\lambda_i}{\sum\limits_{r=1}^{R}\lambda_r} = \frac{\lambda_i}{\lambda}. \label{pi}
\end{align}
So the probability that $S$ takes on the value of $S_i$ is $p_i~\forall i = 1, 2, \cdots, R$. Therefore the probability distribution function (pdf) of $S$ is given by
\begin{equation}
f_S(s) = \sum\limits_{r=1}^R{p_r f_{S_r}(s)}.
\label{eq:pdf_S}
\end{equation}
Then the mean and the second moment of $S$ are simply
\begin{align}
\textnormal{E}[S] = \sum\limits_{r=1}^R{p_r \textnormal{E}[S_r]},~\textnormal{E}[S^2] = \sum\limits_{r=1}^R{p_r \textnormal{E}[S_r^2]}.
\label{eq:meanVarS}
\end{align}
Using \eqref{pi} and \eqref{eq:meanVarS} in \eqref{eq:timeFCFS}, we obtain,
\begin{align}
T_{\textnormal{FCFS}}^i = \textnormal{E}[S_i]+ \frac{\sum\limits_{r=1}^R \lambda_r \left[\textnormal{V}[S_r]+\textnormal{E}[S_r]^2\right]}{2\left(1-\sum\limits_{r=1}^R  \lambda_r \textnormal{E}[S_r]\right)}. \label{fcfs_ub1} 
\end{align}

\subsection{Latency Analysis of Homogenous DSS} \label{homoSystem}
\begin{figure}%
\centering
\includegraphics[scale = 0.35]{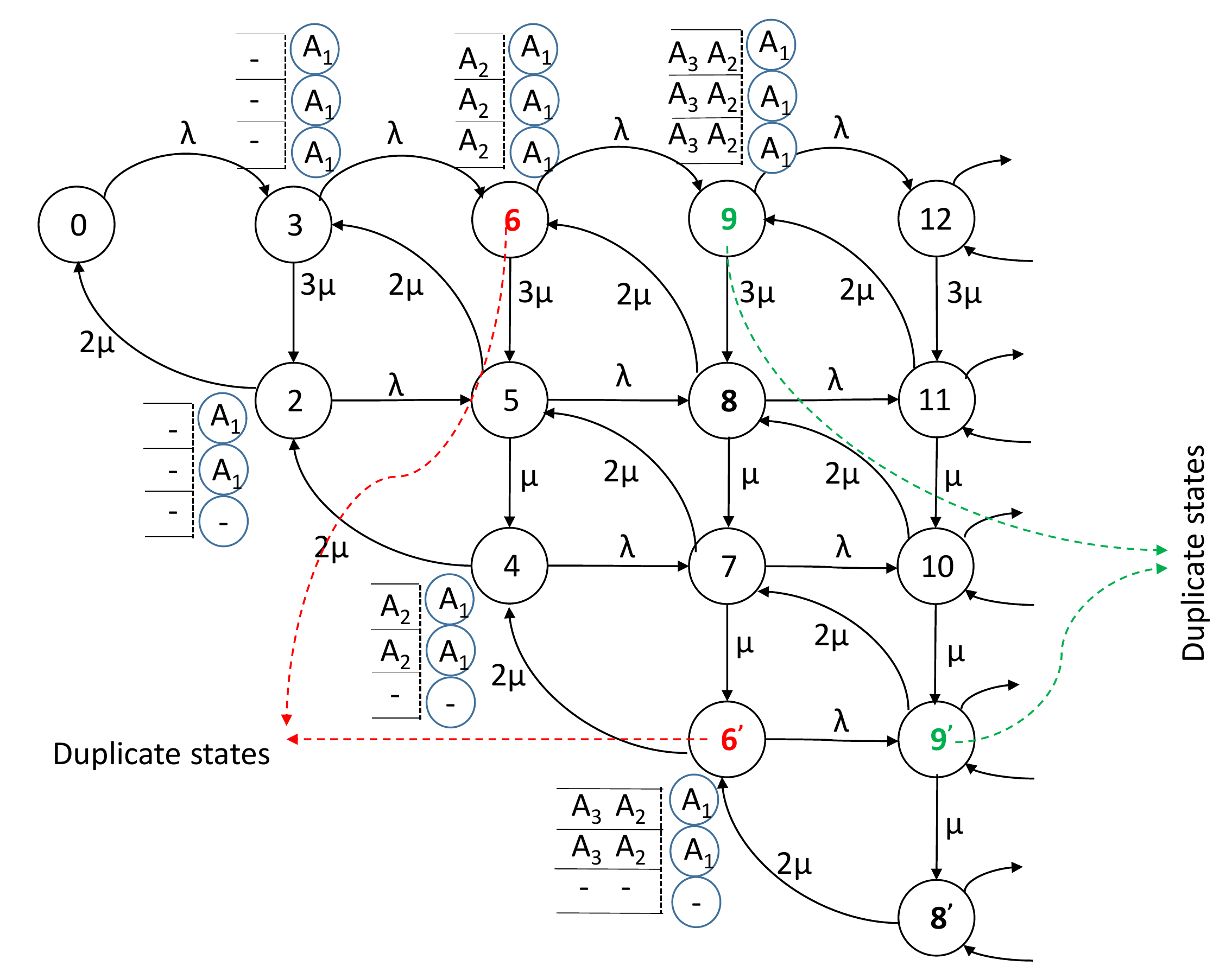}%
\caption{Markov chain for a $(3,2)$ Fork-Join system.}%
\label{markovChain}%
\end{figure}

An exact latency analysis of the $(n,k)$ DSS is prohibitively complex because the Markov chain has a state space with infinite states in at least $k$ dimensions. This is exemplified in Figure~\ref{markovChain} which shows the Markov chain evolution for a $(3,2)$ DSS. Each state is characterized by the number of jobs in the system. The arrival and service rates of jobs are $\lambda$ and $\mu$ respectively. We note that as more jobs enter the system, the Markov Chain starts growing in two-dimensions and results in multiple states with the same number of jobs in the system such as states $6$ and $6^{'}$. Thus, we note that an exact analysis of the F-J system is very complex. Therefore, we review existing upper- and lower-bounds for the average latency of homogenous DSS.  

\subsubsection{Lower Bound on Average Latency}
\label{sec:lowerBound}
In a $(n,k)$ DSS, a job is considered finished when $k$ out of $n$ servers finish that job. This is equivalent to each job going through $k$ stages sequentially, where the transition from one stage to the next occurs when one of the remaining servers finishes a sub-task of the job \cite{varki}. We note that at any stage $s$, the maximum possible service rate for a job that is not finished yet is $(n-s+1)\mu^{'}$, where $\mu^{'} = \frac{f k\mu}{l}$. This happens when all the remaining sub-tasks of a job are at the head of their queues. Thus, we can enhance the latency performance in each stage $s$ by approximating it with a M/M/1 system with service rate $(n-s+1)\mu^{'}$. Then, the average latency of the original system (denoted by $T$), can be lower bounded as
\begin{equation}
T \geq T_{\textnormal{LB}} = \sum\limits_{i=1}^k{\frac{1}{(n-i+1)\mu^{'}-\lambda}},
\label{eq:lbFJ}
\end{equation} 
where $T_{\textnormal{LB}}$ denotes the lower bound on the average latency of the F-J system.

\subsubsection{Upper Bound on Average Latency}
\label{sec:upperBound}
To upper-bound the performance of the $(n,k)$ F-J system, we degrade its performance by approximating it with a $(n,k)$ Split-Merge (SM) system, proposed in \cite{gauri13}. In the $(n,k)$ SM system, after a server finishes a copy of a job, it is blocked and not allowed to accept new jobs until all $k$ copies of the current job are finished. When $k$ copies of a job are finished, the copies of that job at remaining $n-k$ servers exit the system immediately. The SM system thus can be modeled as a M/G/1 system with arrival rate $\lambda$ and a service distribution that follows $k^{\text{th}}$ order statistics \cite{ross02} and is described here for reference. 

Let $X_1$, $X_2$, ..., $X_n$ be $n$ i.i.d random variables (rv). Now if we order the rv's in ascending order to get, $X_{1,n}<X_{2,n}\cdots<X_{k,n}\cdots<X_{n,n}$, then the distribution of the $k^{\text{th}}$ smallest value, $X_{k,n}$, is called the $k^{\textnormal{th}}$ order statistics. The pdf of $X_{k,n}$ is given by\footnote{ The result in \eqref{eq:pdf_kthOrd} can be understood as follows. First select groups of $k-1$, $1$, and $n-k$ servers out of $n$ servers in $\binom{n}{k-1,1,n-k}$ possible ways. Then the pdf of service time for the singled-out server is simply $f_X(x)$. Now since $X_i$ are i.i.d random variables, the probability that the selected $k-1$ servers finish their jobs before the singled-out server is $F_X{(x)}^{k-1}$. Similarly, the probability that $n-k$ servers finish their jobs after the singled-out server is ${(1-F_X(x))}^{n-k}$.}
\begin{equation}
f_{X_{k,n}}(x) = \binom{n}{k-1,1,n-k}F_X{(x)}^{k-1}{(1-F_X(x))}^{n-k}f_X(x)
\label{eq:pdf_kthOrd}
\end{equation}
where $F_X(x)$ and $f_X(x)$ are the cumulative density function and pdf of $X_i$ respectively for all $i$. The average latency of the F-J system is thus upper-bounded by the average latency for the SM system, $T_{\textnormal{SM}}$ as,
\begin{eqnarray}
T \leq T_{\textnormal{SM}} = \underbrace{\text{E}[X_{k,n}]}_\text{service time}+ \underbrace{\frac{\lambda \left[\textnormal{V}[X_{k,n}]+\textnormal{E}[X_{k,n}]^2\right]}{2(1-\lambda \textnormal{E}[X_{k,n}]))}}_\text{waiting time},
\label{eq:timeMG1}
\end{eqnarray}
where the average service time is simply the expectation, $\text{E}[X_{k,n}]$ and the average waiting time for a M/G/1 system given by the P-K formula in \eqref{waitFCFS}. Now if $X_i$ is exponential with mean $1/\mu^{'}$ (where $\mu^{'} = \frac{f k\mu}{l}$), then the mean and variance of $X_{k,n}$ are given by,
\begin{equation}
\textnormal{E}[X_{k,n}] = \frac{H_{n-k,n}^1}{\mu^{'}}, \textnormal{V}[X_{k,n}] = \frac{H_{n-k,n}^2}{{\mu^{'}}^2},
\label{eq:kthOrd}
\end{equation}
where $H_{x,y}^z$ is a generalized harmonic number of order $z$ defined by
\begin{equation}
H_{x,y}^z = \sum\limits_{j=x+1}^y \frac{1}{j^z},
\label{eq:harmMean}
\end{equation}
for some positive integers $x, y$ and $z$.

\section{Main Results}
\label{meanRespTime}
Section~\ref{homoSystem} presented bounds on the average latency for the $(n,k)$ F-J system. To extend the lower-bound result \eqref{eq:lbFJ} to a heterogeneous FJ system, a naive approach would be to approximate it with a homogenous FJ system with jobs of class $i$ only while evaluating the lower bound on average latency of class $i$. Thus a naive lower-bound on the average latency for jobs of class $i$ is, 
\begin{align}
\label{naiveBound}
T_{\textnormal{naive}}^i \geq \sum\limits_{j=0}^{k_i-1} \frac{1}{(n-j)\mu_i-\lambda_i}.
\end{align}
This lower bound holds true irrespective of the scheduling policy used in the heterogeneous system. However, this is a loose bound as it ignores the dependency of response time for a job of class $i$ on the jobs of other classes in the system which compete for the service at the same server. 

Therefore, through a rigorous latency analysis of various scheduling policies, we next account for this inter-dependency in average latency of different classes and present lower and upper bounds for the heterogeneous FJ system. To this end, we first define a set of variables for a compact presentation of the results. The operational meaning of these variables will become clear when we present the proof of the results.
\begin{itemize}
\item $(n, k_i)$ is the MDS code used to store data of class $i$.
\item $l_i$ is the file-size for class $i$.
\item $\lambda_i$ is the arrival rate for jobs of class $i$.
\item $\mu_i = k_i f \mu/l_i$ is the effective service rate for jobs of class $i$, where $\mu$ is the service rate per unit file size.
\item $\rho_i = \frac{\lambda_i}{\mu_i}$ is the server utilization factor for class $i$.
\item $\mathcal{S}_i = \sum\limits_{r=1}^i \rho_r H_{n-k_r,n}^1$.
\end{itemize}

\subsection{Main Results}
Lemma~\ref{Lemma1} gives the stability conditions of the heterogeneous DSS for various scheduling policies. The upper- and lower-bounds on the average latency for various scheduling policies are presented in Theorem~\ref{Theorem1} and \ref{Theorem2} respectively.
\label{mainResult}
\begin{Lem}\label{Lemma1}
For a $(n, k_1, k_2, ..., k_R)$ Fork-Join system to be stable, the following condition must be satisfied at each node.
%\vspace{-30pt}
\begin{itemize}
\item FCFS scheduling
\begin{align}
\left(\sum\limits_{r=1}^R{k_r \lambda_r}\right)\left(\sum\limits_{r=1}^R{\frac{\lambda_r l_r}{k_r}}\right) < nf\mu \sum\limits_{r=1}^R{\lambda_r}.
\end{align}
\item Preemptive/Non-preemptive priority scheduling
\begin{align}
\sum\limits_{r=1}^R{\lambda_r l_r} < nf\mu.
\end{align}
\end{itemize}
\end{Lem}

Next, to upper-bound the average latency, we extend the Split-Merge (SM) system (defined in Section~\ref{sec:upperBound}) to $R$ data classes, keeping the scheduling policy same as that for the original system. Then for a given scheduling policy, the upper-bound on average latency is basically the average latency of the corresponding SM system. This in turn is sum of the average service time and waiting time which can be determined by noting the equivalence between the SM system as a M/G/1 system as described in Section~\ref{sec:upperBound}. We thus obtain the following upper-bounds on the average latency for different scheduling policies.
\begin{Thrm}\label{Theorem1} 
The average latency for job requests of class $i$ in a $(n, k_1, k_2, ..., k_R)$ Fork-Join system is upper-bounded as follows:
\begin{itemize}
\item FCFS scheduling
\begin{align}
\label{eq:fcfs_ub}
T_{\textnormal{FCFS}}^{i} \leq \underbrace{\frac{H_{n-k_i,n}^1}{\mu_i}}_\text{Service time}+ \underbrace{\frac{\sum\limits_{r=1}^R{\lambda_r[H_{n-k_r,n}^2 + {(H_{n-k_r,n}^1)}^2]/ {\mu_r}^2}}{2\left(1-\mathcal{S}_R\right)}}_\text{Waiting time}.
\end{align}
The bound is valid only when $\mathcal{S}_R < 1$.
\item  Non-preemptive priority scheduling\footnote{Without loss of generality, we set the classes in the order of decreasing priority as $1>2>\cdots>R$.}
\begin{align}
\label{npq_ub}
T^i_{\textnormal{N-PQ}} \leq \frac{H_{n-k_i,n}^1}{\mu_i} + \frac{\sum\limits_{r=1}^R{\lambda_r[H_{n-k_r,n}^2  + {(H_{n-k_r,n}^1)}^2]/{\mu_r}^2}}{2\left(1-\mathcal{S}_{i-1}\right)\left(1-\mathcal{S}_i\right)}.
\end{align}
The bound is valid only when $\mathcal{S}_i < 1$.
\item Preemptive priority scheduling\footnotemark[\value{footnote}]
\begin{align}
\label{pq_ub}
T^i_{\textnormal{PQ}} \leq \frac{H_{n-k_i,n}^1}{\mu_i \left(1-\mathcal{S}_{i-1}\right)} + \frac{\sum\limits_{r=1}^i \lambda_r[H_{n-k_r,n}^2  + {(H_{n-k_r,n}^1)}^2]/{\mu_r}^2}{2\left(1-\mathcal{S}_{i-1}\right)\left(1-\mathcal{S}_i\right)}.
\end{align}
The bound is valid only when $\mathcal{S}_i < 1$.
\end{itemize}
\end{Thrm} 

We now define an additional set of variables for compact presentation of the results in Theorem 2.
\begin{itemize}
\item Without loss of generality, assume the classes are relabeled such that $k_1 \le k_2 \le ... \le k_R$. 
Then for class $i$, we define $c_s$ as,
\begin{align}
c_s = \begin{cases} 0, &1\leq s \leq k_1\\ 1, &k_1 < s \leq k_2\\\vdots\\i-1, &k_{i-1} < s \leq k_i \end{cases}. \label{csdef}
\end{align}
\item At a stage $s$, let $\mathcal{R}_s^{i}$ denote the set of classes with priority higher than class $i$ and that have not been finished yet.
\vspace{3 pt}
\item $t_{s,i} = \frac{\lambda_i}{(n-s+1)\mu_i}$ at stage $s$ and class $i$.
\vspace{4 pt}
\item $\mathcal{Z}_s^i = 1-\sum\limits_{r\in \mathcal{R}_{s}^{i}} t_{s,r}$ at stage $s$ and class $i$.
\end{itemize}

For obtaining a lower-bound on the average latency, we enhance the performance of the original system similar to the process described in Section~\ref{sec:upperBound}. The processing of a job of class $i$ is modeled as completing $k_i$ sequential stages (or sub-tasks). Then we enhance the latency performance for job of class $i$ in stage $s$ by assuming the maximum possible service rate for it, i.e, $(n-s+1)\mu_i$. However, at stage $s$, there may also be unfinished sub-tasks of jobs of other classes which can be served with maximum possible service rate of $(n-s+1)\mu_j$, where $j\neq i$. Due to this, we model the performance of each enhanced stage as a M/G/1 system. We thus obtain the following lower-bounds on the average latency for different scheduling policies.

\begin{Thrm}\label{Theorem2} 
The average latency for job requests of class $i$ in a $(n, k_1, k_2, ..., k_R)$ Fork-Join system is lower-bounded as follows:
\begin{itemize}
\item FCFS scheduling
\begin{equation}
\label{eq:fcfs_lb}
T_{\textnormal{FCFS}}^i \geq \sum\limits_{s=1}^{k_i} \left(\underbrace{\frac{t_{s,i}}{\lambda_i}}_\text{service time}+ \underbrace{\frac{\sum\limits_{r=c_s+1}^R \frac{t_{s,r}^2}{\lambda_r}}{1-\sum\limits_{r=c_s+1}^{R}t_{s,r}}}_\text{waiting time}\right).
\end{equation}
\item Non-Preemptive priority scheduling\footnotemark[\value{footnote}]
\begin{align}
\label{npq_lbp}
T^i_{\textnormal{N-PQ}} \geq  \sum\limits_{s=1}^{k_i} \left(\frac{t_{s,i}}{\lambda_i} + \frac{\sum\limits_{r=c_s+1}^R \frac{t_{s,r}^2}{\lambda_r}}{\mathcal{Z}_s^i\left(\mathcal{Z}_s^i-t_{s,i}\right)}\right).
\end{align}
\item Preemptive priority scheduling\footnotemark[\value{footnote}]
\begin{align}
\label{pq_lbp}
T^i_{\textnormal{PQ}} \geq  \sum\limits_{s=1}^{k_i} \left(\frac{t_{s,i}}{\lambda_i \mathcal{Z}_s^i} + \frac{\sum\limits_{r\in\mathcal{R}_s^i\cup{i}}\frac{t_{s,r}^2}{\lambda_r}}{\mathcal{Z}_s^i\left(\mathcal{Z}_s^i-t_{s,i}\right)}\right).
\end{align}
\end{itemize}
\end{Thrm}

\subsection{Proofs for FCFS scheduling}
\label{proof}
We now present the proofs for the stability condition and the bounds on average latency for the FCFS scheduling policy. The proofs for the remaining results are given in Appendix~\ref{append1}.

\subsubsection{Proof of Lemma 1-FCFS scheduling} Consider any server in the $(n, k_1, k_2, ..., k_R)$ Fork-Join system. Jobs of class $r$ enter the queue with rate $\lambda_r$. Each new job of class $r$ exits the system when $k_r$ sub-tasks of that job are completed. The remaining $n-k_r$ sub-tasks are then cleared from the system. Thus for each job of class $r$, $\frac{(n-k_r)}{n}$ fraction of the sub-tasks are deleted and hence the effective arrival rate of jobs of class $r$ at any server is $\lambda_r\left(1-\frac{n-k_r}{n}\right) = \frac{k_r \lambda_r}{n}$. Thus the overall arrival rate at any server, $\lambda_{\textnormal{eff}}$, is 
\begin{align}
\lambda_{\textnormal{eff}} = \sum\limits_{r=1}^R{\frac{k_r \lambda_r}{n}}.
\label{lambda_eff}
\end{align} 
Let $S$ denote the service distribution for a single-server FCFS system serving $R$ data classes. Then from \eqref{eq:meanVarS}, the mean service time at a server is 
\begin{align}
E[S] = \sum\limits_{r=1}^R{p_r E[S_r]} = \sum\limits_{r=1}^R{\frac{\lambda_r}{\mu_r \sum\limits_{r=1}^{R}\lambda_r}},	 
		\label{meanServiceTime}	
\end{align}	
where \eqref{meanServiceTime} follows from \eqref{pi} and the assumption that the service time for a job of class $i$ is exponential with rate $\mu_r$. To ensure stability, the net arrival rate should be less than the average service rate at each server. Thus from \eqref{lambda_eff} and \eqref{meanServiceTime} the stability condition of each queue is 
\begin{align}
\sum\limits_{r=1}^R{\frac{k_r \lambda_r}{n}} < {\left(\sum\limits_{r=1}^R \frac{\lambda_r}{\mu_r \sum\limits_{r=1}^{R}\lambda_r}\right)}^{-1}, \nonumber \end{align}
Since $\mu_r = \frac{f k_r \mu}{l_r}$ and the term $\sum\limits_{r=1}^{R}\lambda_r$ is a constant, with simple algebraic manipulations we arrive at
\begin{align}
\left(\sum\limits_{r=1}^R{k_r \lambda_r}\right)\left(\sum\limits_{r=1}^R{\frac{\lambda_r l_r}{k_r}}\right) < nf\mu \sum\limits_{r=1}^R{\lambda_r}.
\end{align}
This completes the proof of stability condition for FCFS scheduling.
%%%%%%%%%%%%%%%%%%%%%%%%%%%%%%%%%%%%%%%%%%%%%%%%%%%%%%%%%%%%%% 
\subsubsection{Proof of Theorem 1-FCFS scheduling}\label{upFCFS}
The FCFS system can be modeled as a M/G/1 queuing system with arrival rate $\lambda = \sum\limits_{r=1}^R \lambda_r$ and a general service time distribution $S$. Then the average latency for a job of class $i$ in a FCFS scheduling system is given by \eqref{fcfs_ub1} as,
\begin{align}
T_{\textnormal{fcfs}}^i = \textnormal{E}[S_i]+ \frac{\sum\limits_{r=1}^R \lambda_r \left[\textnormal{V}[S_r]+\textnormal{E}[S_r]^2\right]}{2\left(1-\sum\limits_{r=1}^R  \lambda_r \textnormal{E}[S_r]\right)}. \nonumber 
\end{align}

\begin{figure}%
\centering
\includegraphics[scale=0.51]{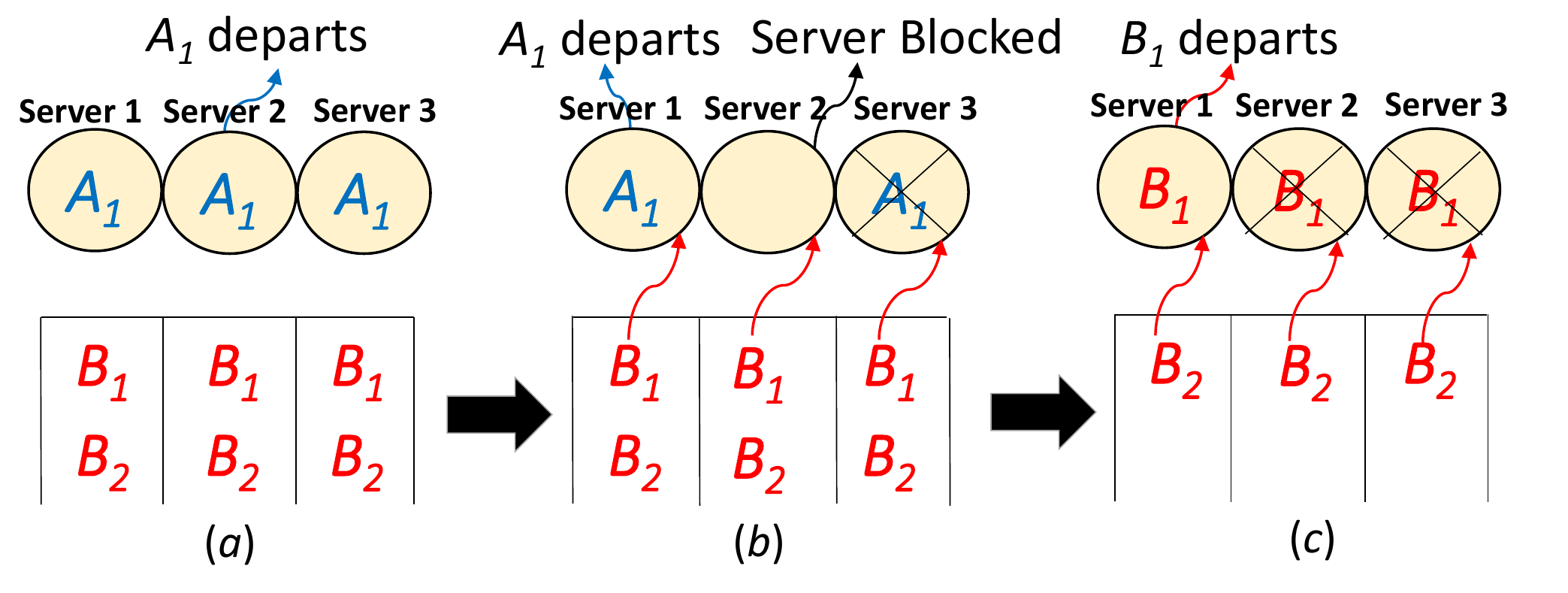}%
\vspace{-7pt}
\caption{Enhanced two-class FJ system with FCFS.}%
\label{blocking}%
\end{figure}

To obtain an upper bound on the average latency, we degrade the FJ system in the following manner. For a job of class $i$, the servers that have finished processing a sub-task of that job are blocked and do not accept new jobs until $k_i$ sub-tasks of that job have been completed. Then the sub-tasks at remaining $n-k_i$ servers exit the system immediately. Fig.~\ref{blocking} illustrates this process using Example~\ref{example}. When $A_1$ is finished at server 2, it is blocked (see Fig.~7(b)) until another $k_A = 2$ copies are finished. Now this performance-degraded system can be modeled as a M/G/1 system where the distribution of the service process, $S_i$, follows $k_i^{\textnormal{th}}$ ordered statistics as described in Section~\ref{sec:upperBound}. Now for any class $i$, the service time at each of the $n$ servers is exponential with mean $1/\mu_i$. Hence from \eqref{eq:kthOrd}, the mean and variance of $S_i$ are,
\begin{align}
\textnormal{E}[S_i] = \frac{H_{n-k_i,n}^1}{\mu_i},~\textnormal{V}[S_i] = \frac{H^2_{n-k_i,n}}{\mu_i^2}. \label{eq:meanVarSi}
\end{align}
Substituting \eqref{eq:meanVarSi} in \eqref{fcfs_ub1}, we get the following upper bound on average latency:
\begin{align}
T_{\textnormal{FCFS}}^{i} \leq  \underbrace{\frac{H_{n-k_i,n}^1}{\mu_i}}_\text{service time} +  \underbrace{\frac{\sum\limits_{r=1}^R{\lambda_r[H_{n-k_r,n}^2 + {(H_{n-k_r,n}^1)}^2]/ {\mu_r}^2}}{2\left(1-\mathcal{S}_R\right)}}_\text{waiting time},
\end{align}
where $\mathcal{S}_R = \sum\limits_{r=1}^R \rho_r H_{n-k_r,n}^1$ and $\rho_r = \lambda_r/\mu_r$.
This concludes the proof of upper bound on the average latency for FCFS scheduling.

%%%%%%%%%%%%%%%%%%%%%%%%%%%%%%%%%%%%%%%%%%%%%%%%%%%%%%%%%%%%%%%%
\subsubsection{Proof of Theorem 2-FCFS scheduling}\label{lowFCFS}
For the purpose of obtaining a lower bound on the average latency of class $i$, using insights from Section~\ref{sec:lowerBound}, we map the parallel processing in the proposed FJ system to a sequential process consisting of $k_i$ processing stages for $k_i$ sub-tasks of a job of class $i$. The transition from one stage to the next occurs when one of the remaining servers finishes a sub-task of the job. Let $c_s$ denotes the number of classes that are finished before start of stage $s$, defined in \eqref{csdef}. The processing in each stage $s$ corresponds to a single-server FCFS system with jobs of all but classes $1, 2, \cdots, c_s$. Then, using \eqref{eq:timeFCFS} for the FCFS sub-system at stage $s$, the average latency for a sub-task of a job of class $i$ in stage $s$ is given by, 
\begin{align}
T_{\textnormal{FCFS},s}^i &= \textnormal{E}[S_i^s]+ \frac{\lambda \textnormal{E}[{(S^s)}^2]}{2(1-\lambda \textnormal{E}[S^s]))}, \label{tmp}
\end{align} 
where $S^s$ is a r.v. denoting the service time for any sub-task in stage $s$ and $S_i^s$ denotes the service time for a sub-task of class $i$ in stage $s$. Now the moments of $S^s$ and $S_i^s$ are related to each other in the same way as the moments of $S$ and $S_i$ in \eqref{eq:meanVarS}. So we have,
\begin{align}
\textnormal{E}[S^s] = \sum\limits_{r=c_s+1}^R{p_r \textnormal{E}[S_r^s]},~~\textnormal{E}[{(S^s)}^2] = \sum\limits_{r=c_s+1}^R{p_r \textnormal{E}[{(S_r^s)}^2]}.
\label{meanVarSs}
\end{align}
Substituting \eqref{meanVarSs} in \eqref{tmp}, we get
\begin{align}
T_{\textnormal{FCFS},s,c_s}^i &=\textnormal{E}[S_i^s] + \frac{\sum\limits_{r=c_s+1}^R \lambda_r \textnormal{E}[{(S^s_i)}^2]}{2\left(1-\sum\limits_{r=c_s+1}^R \lambda_r \textnormal{E}[S^s_i]\right)}.
\end{align}
Now we note that at any stage $s$, the maximum possible service rate for a job of class $j$ that is not finished yet is $(n-s+1)\mu_j$. This happens when all the remaining sub-tasks of job of class $j$ are at the head of their buffers. Thus, we can enhance the latency performance in each stage $s$ by approximating it with a M/G/1 system with service rate $(n-s+1)\mu_j$ for jobs of class $j$. Then, the average latency for sub-task of job of class $i$ in stage $s$ is lower bounded as,
\begin{align}
T_{\textnormal{FCFS},s,c_s}^i  &\geq \frac{1}{(n-s+1)\mu_i} + \frac{\sum\limits_{r=c_s+1}^R{\frac{\lambda_r}{(n-s+1)\mu_r^2}}}{1- \sum\limits_{r=c_s+1}^R{\frac{\lambda_r}{(n-s+1)\mu_r}}}, \label{latEnhcd}
\end{align} 
Finally, the average latency for class $i$ in this enhanced system is simply $\sum\limits_{s=1}^{k_i} T_{\textnormal{FCFS},s,c_s}^i$. This gives us
\begin{align}
T_{\textnormal{FCFS}}^i \geq \underbrace{\sum\limits_{s=1}^{k_i} \frac{t_{s,i}}{\lambda_i}}_\text{service time} + \underbrace{\sum\limits_{s=1}^{k_i}\left(\frac{\sum\limits_{r=c_s+1}^R \frac{t_{s,r}}{(n-s+1)\mu_r}}{1-\sum\limits_{r=c_s+1}^{R}t_{s,r}}\right)}_\text{waiting time}, \nonumber
\end{align}
where $t_{s,i} = \frac{\lambda_i}{(n-s+1)\mu_i}$.
This concludes the proof of lower bound on the average latency for FCFS scheduling.

%%%%%%%%%%%%%%%%%%%%%%%%%%%%%%%%%%%%%%%%%%%%%%%%%%%
\section{Quantitative Results and Discussion}
\label{result}
In this section, we use Monte-Carlo simulations of a heterogeneous Fork-Join system to study the impact of varying various system parameters on the average latency of different classes and the energy efficiency of DSS. For simplicity, the number of data classes is set to $2$. Data of class $1$ is stored using $(n,k_1)=(10,5)$ MDS code. Data of class $2$ is stored using $(10, k_2)$ MDS code where $k_2$ is varied from $1$ to $10$. Arrival rates for the two classes are set as: $\lambda_1 = 0.15$ and $\lambda_2=0.5$. The job size for both the classes is set to 1 kilobits. Job requests for both the classes are served using full redundancy (i.e., $r_1 = r_2 = n$). We set the power consumption parameters by using the data for Intel Xeon family of CPUs\footnote{We use the power consumption parameters of ``Deeper Sleep'' state for our low-power state.} and associated platform components \cite{Liu14}. Therefore, we set $C_0=203.13$~W, $P_a=120$~W, $C_l=15$~W, $w_l=6$~s, and $P_l=13.1$~W. The CPU frequency, $f$ is set to $1$ unless mentioned otherwise.

\subsection{Impact of fault-tolerance and service rate}\label{faultTol}
The behavior of average latency with respect to change in fault-tolerance $k$, is governed by two opposing factors.
\begin{itemize}
\item Increasing $k$ reduces the number of servers available for serving the next job in queue, thus resulting in an increase in latency.
\item Increasing $k$ increases the effective service rate ($k\mu$) of each server as each server stores a smaller fraction ($\frac{l}{k}$) of the job. This decreases the average latency.
\end{itemize}
Fig.~\ref{muLow_class2} shows the average latency for jobs of class $2$ versus $k_2$ for the FCFS system with $\mu= 1/6$ and $1$\footnote{In our work, we set file size to be multiples of 1 kilobits. So $\mu$ is defined per kilobit.}. The file size for both classes are equal to $1$~kb. We note that the average latency increases on increasing $k_2$. This is because $\mu$ is large enough, so the increment in latency due to the first factor dominates the decrease in latency due to the second factor. We also note that the bounds are somewhat loose at high values of $k_2$ and low values of $\mu$. In particular, the lower bound becomes loose because at each processing stage of the serial Fork-Join system, the difference between the actual service rate and its bound at $s^{\textnormal{th}}$ stage of processing (i.e.\ $(n-s+1)\mu_i$) for jobs of class $i$ increases with increase in $k$ and decrease in $\mu$. Similarly the upper bound becomes worse because the service time lost due to blocking increases significantly at low $\mu$ and high $k$ values. This is because the remaining sub-tasks are served really slow (low $\mu$) and the blocking continues until a large number of sub-tasks (high $k$) are finished. Finally, as expected, we note that the naive lower bound on latency of class 2 is loose as compared to the proposed lower bound for the FCFS system.

\subsection{Impact of coding on energy efficiency and storage space}\label{ee_bandwidth}
\begin{figure}[!ht]
\centering
\includegraphics[scale=0.33]{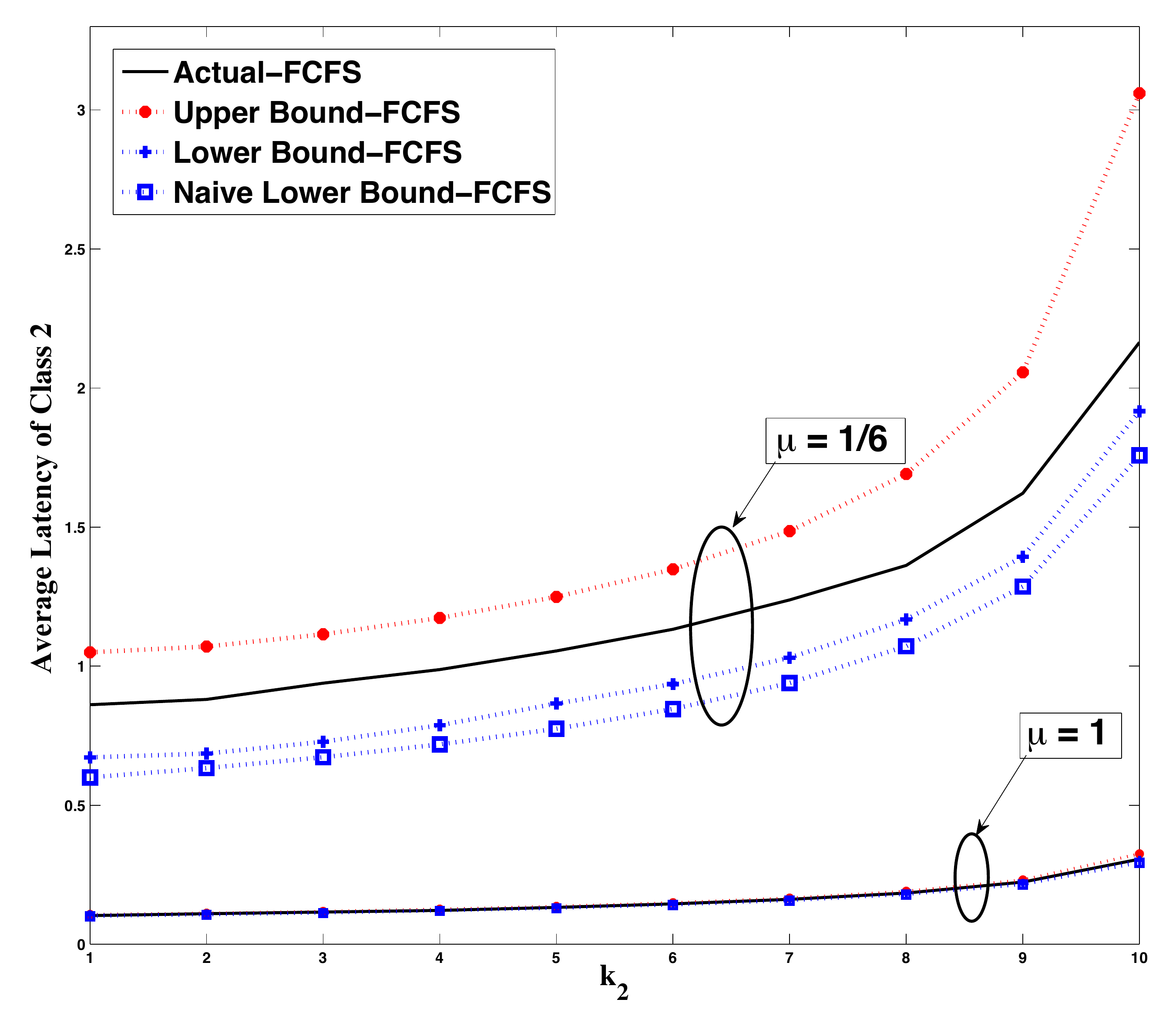}
\vspace{-5pt}
\caption{\footnotesize{Latency of a data-class increases with increase in its code-rate and decreases with increase in service rate.}}
\vspace{0pt}
\label{muLow_class2}
\end{figure}

\begin{figure}[!ht]
\centering
\includegraphics[scale=0.50]{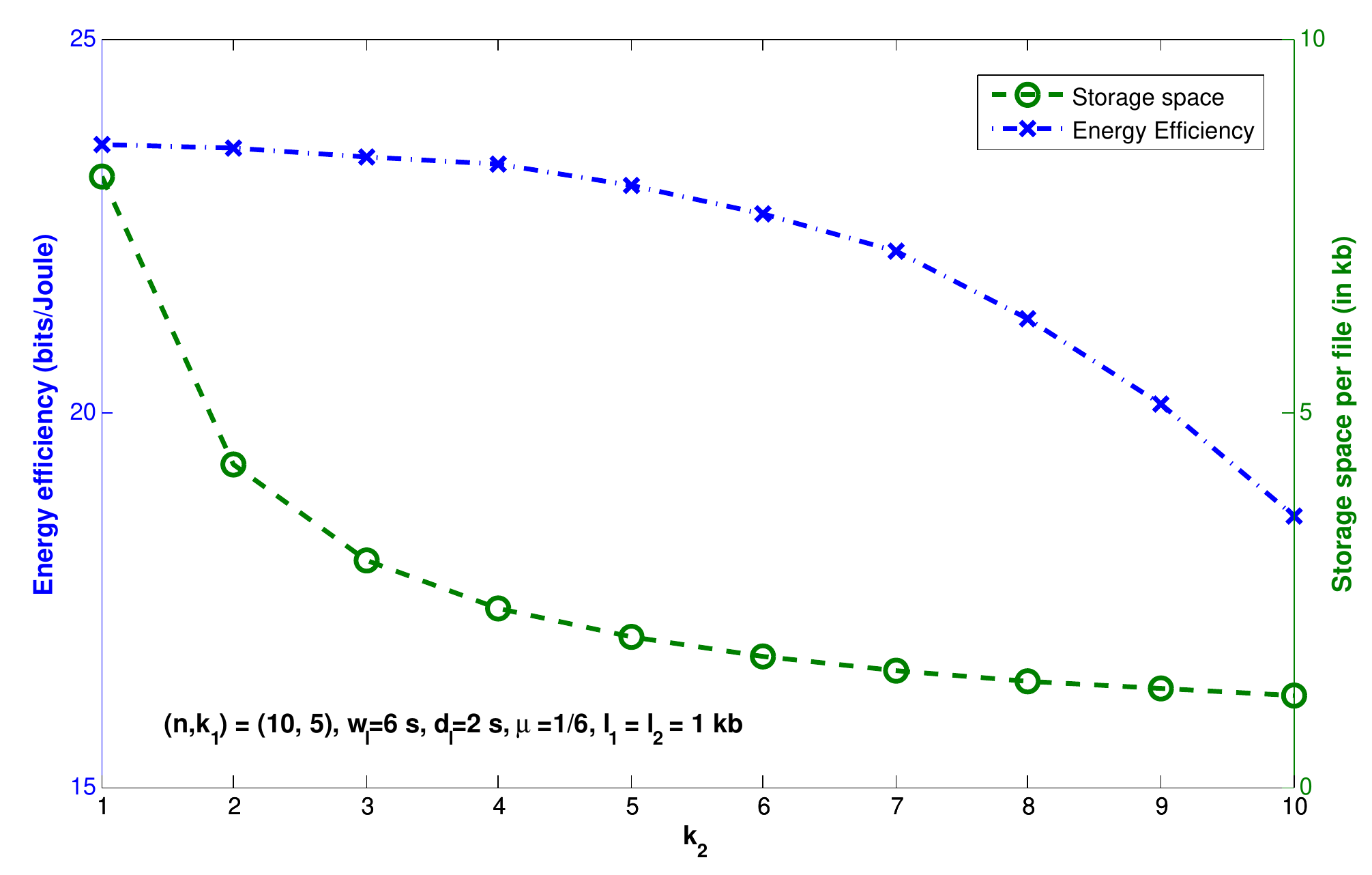}
\vspace{-5pt}
\caption{\footnotesize{Tradeoff between energy efficiency of DSS and storage space per file with variation in code-rate.}}
\vspace{-4pt}
\label{ee_nwbw}
\end{figure}

Fig.~\ref{ee_nwbw} illustrates the impact of varying the code rate ($k/n~\text{for}~(n,k)~\text{MDS code}$) on the latency, energy efficiency and network bandwidth of system. At one extreme is the $(n,1)$ replication code with code rate $1/n$ that has minimum latency (see Fig.~\ref{muLow_class2}) and maximum energy efficiency. This is because we just wait for any one server to finish the job. For a fixed $n$, low latency translates to higher data throughput and hence higher energy efficiency. However, the total storage space per file for $(n,1)$ code is $nl$ where $l$ is file size. Hence the (write) network bandwidth is maximum at $k=1$. At the other extreme is $(n,n)$ code with no fault-tolerance and no storage overhead (storage size is $l$). But it suffers from high latency and low energy efficiency. This is because we need to wait for all the servers to finish their sub-tasks of a job before the job is completed. Hence latency and throughput suffers which in turn decreases energy efficiency.		
%highlights the impact of varying $k$ (hence the code rate $k/n$), on the energy efficiency and storage space required per file using FCFS scheduling policy. Since increasing $k$ increases the average latency (see Fig.~\ref{muLow_class2}), the data throughput decreases, hence the energy efficiency decreases. This corroborates our previous result (in Section~\ref{sysModel} for single server, single class system) that the energy efficiency of DSS is inversely proportional to its latency. The required storage space per file (and hence the network bandwidth) also decreases on increasing the code rate.

\subsection{Impact of number of servers in DSS}\label{volume_DSS}
\begin{figure}[!ht]
\centering
\includegraphics[scale=0.50]{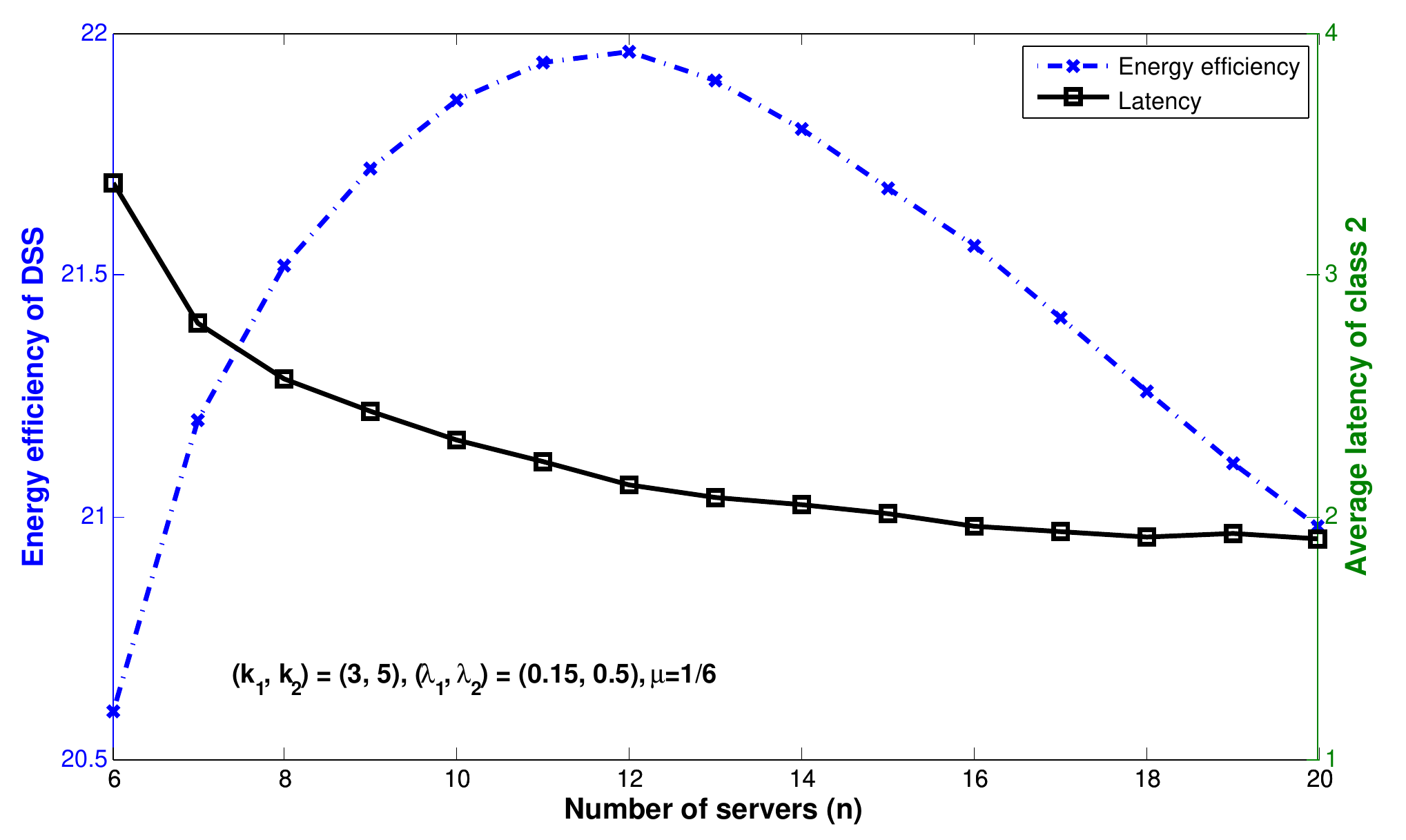}
\vspace{-5pt}
\caption{\footnotesize{Energy efficiency increases and attains a maxima as number of servers is increased while latency behaves in an inverse fashion.}}
\vspace{-4pt}
\label{size_DSS}
\end{figure}

Fig.~\ref{size_DSS} shows the impact of increasing the number of servers ($n$) on the latency and energy efficiency of DSS, while all other system parameters are kept constant. We observed that for low values of $n$, increasing $n$ increases the energy efficiency. This is because of more servers available to serve the job which reduces average latency and thus increase the throughput. The increase in throughput due to lower latency outweighs the increase in energy consumption due to higher $n$. Hence the overall effect is that energy efficiency increases. However at high values of $n$, increasing $n$ results in diminishing returns in latency and throughput. This is because latency improvement is limited by effective service rate ($k\mu/l$) and not the number of servers. At very large $n$, the energy consumption becomes quite significant. Therefore, the energy efficiency begins to decrease at large $n$. We thus conclude that \textbf{there is an optimum value of $n$ that maximizes energy efficiency and has near minimal latency}.
%
%We note that on increasing $n$ from $6$ to $20$, latency begins to decreases but with diminishing returns and eventually flattens out. This is because on increasing $n$, there are more servers available to serve a job request. But when $n$ is large, latency improvement is limited by effective service rate ($k\mu/l$) and not the number of servers. As $n$ is increases, the energy efficiency of DSS first increases, reaches maximum and then starts decreasing. This is because of higher data throughput (lower latency) which initially outweighs the increase in energy consumption on increasing $n$. However at larges value of $n$, the increase in energy consumption becomes dominant and lowers the energy efficiency of DSS. We thus conclude that \textbf{there is an optimum value of $n$ that maximizes energy efficiency and has near minimal latency}.

\subsection{Impact of general service time}\label{general_Service}
\begin{figure}[!ht]
\centering
\includegraphics[scale=0.50]{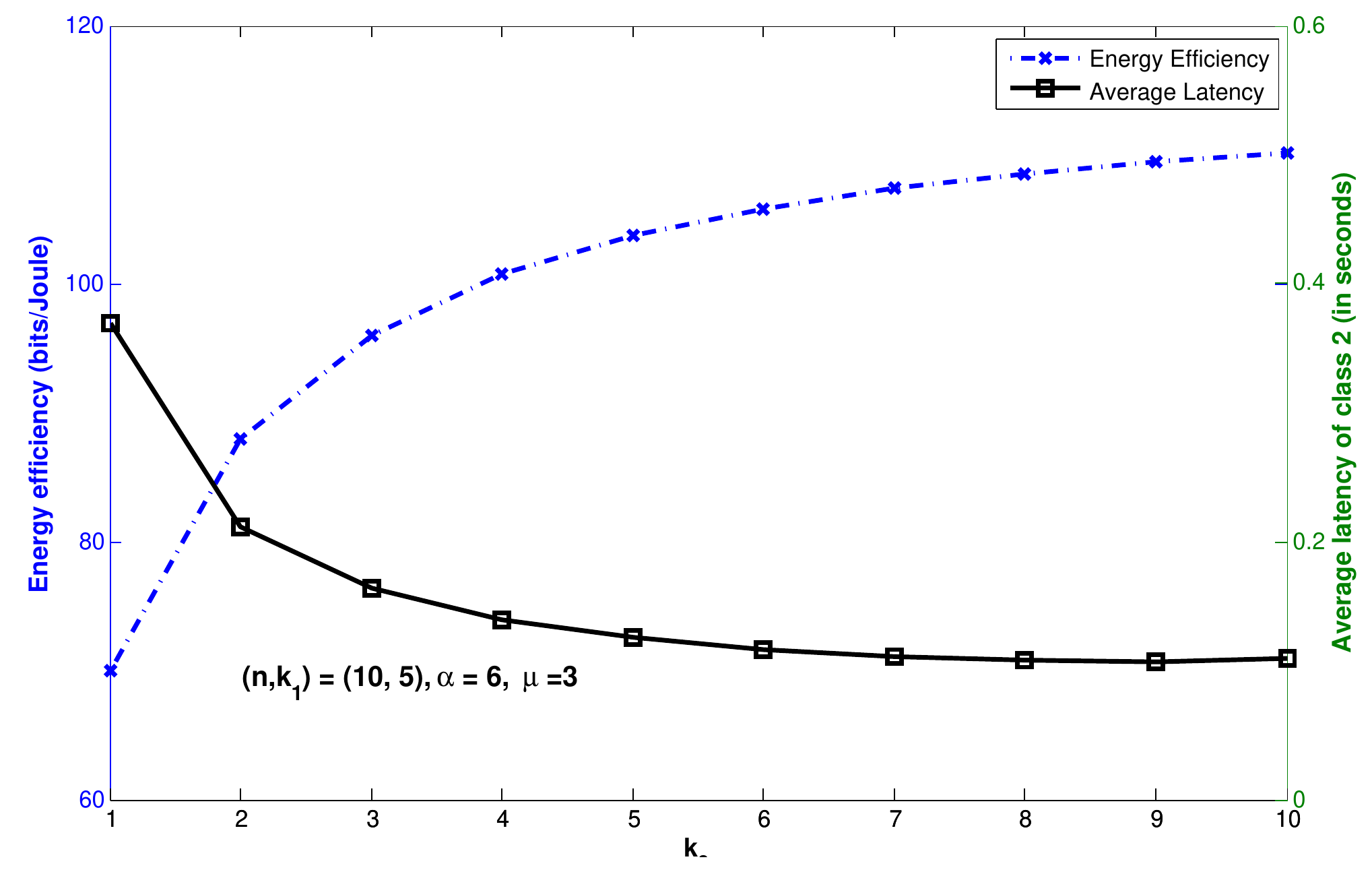}
\vspace{-5pt}
\caption{\footnotesize{A light tailed general service distribution (Pareto-distribution with $\alpha=6$) results in monotonically decreasing latency as a function of code-rate. Energy efficiency follows an inverse behavior.}}
\vspace{-4pt}
\label{pareto_alpha_high}
\end{figure}

\begin{figure}[!ht]
\centering
\includegraphics[scale=0.50]{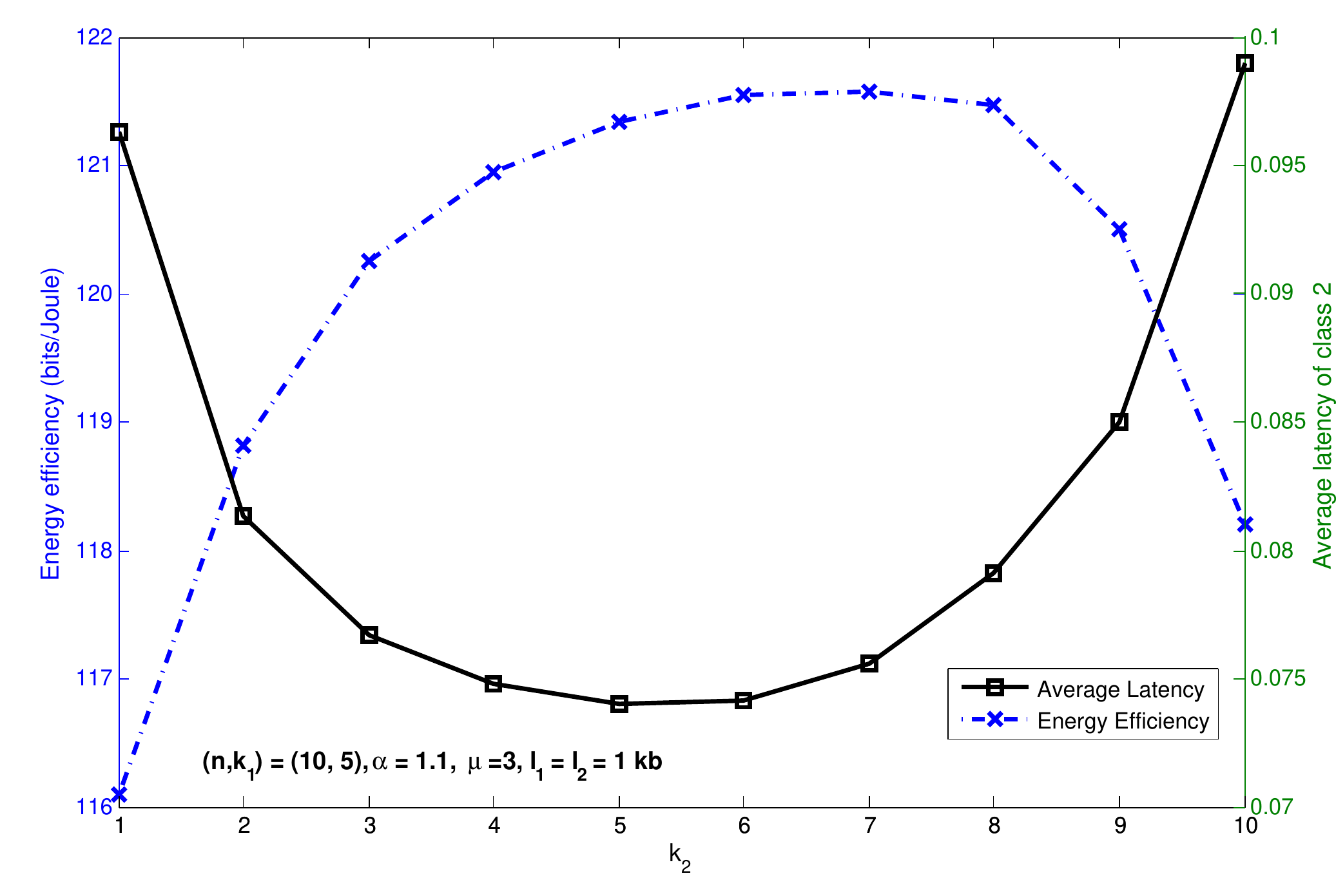}
\vspace{-5pt}
\caption{\footnotesize{A heavy tailed general service distribution (Pareto-distribution with $\alpha=1.1$) results in minimal latency and maximum energy efficiency point as the code-rate is increased.}}
\vspace{-4pt}
\label{pareto_service}
\end{figure}

In most of the practical DSS, the service times are not exponentially distributed but rather have heavy-tail which means that there is a significant probability of very large service times. Pareto distribution has been found to be a good fit for service time distribution in practical DSS \cite{crovella97, Faloutsos99}. Its cumulative distribution function is given by
\begin{equation}
F_S(s) = \begin{cases}
0 & \quad \text{for}~s<s_m \\
1-{\left(\frac{s_m}{s}\right)}^\alpha & \quad \text{for}~s\geq s_m \\
\end{cases}
\label{paretoDist}
\end{equation}
Here $\alpha$ is shape parameter and $x_m$ is the scale parameter. As the value of $\alpha$ decreases the service becomes more heavy-tailed and it becomes infinite for $\alpha \leq 1$. Figures~\ref{pareto_alpha_high} and \ref{pareto_service} show the impact of Pareto service distribution on the latency and energy efficiency of DSS for $\alpha=1.1~\text{and}~6$ respectively. At $\alpha=6$, the service distribution is not very heavy-tailed. So increasing $k_2$ reduces latency of jobs of class 2 due to increase in their effective service rate ($k_2 \mu f/l_2$). However, at $\alpha=1.1$, the service time distribution becomes very heavy-tailed, so as $k_2$ becomes large, the increase in service time due to waiting for more servers (larger $k$) outweighs the decrease due to higher effective service rate. In both cases, we note that latency behaves inversely to the change in latency. We note that as $k_2$ increases from $1$ to $10$, energy efficiency first starts increasing, reaches a maximum  and then starts decreasing for large $k$. We conclude that \textbf{for heavy-tailed service distribution, there exists an optimal code-rate that yield maximum energy efficiency and minimum latency for heavy-tailed service times}.

\subsection{Impact of heavy-tailed arrival distribution}\label{general_arrival}
\begin{figure}[!ht]
\centering
\includegraphics[scale = 0.50]{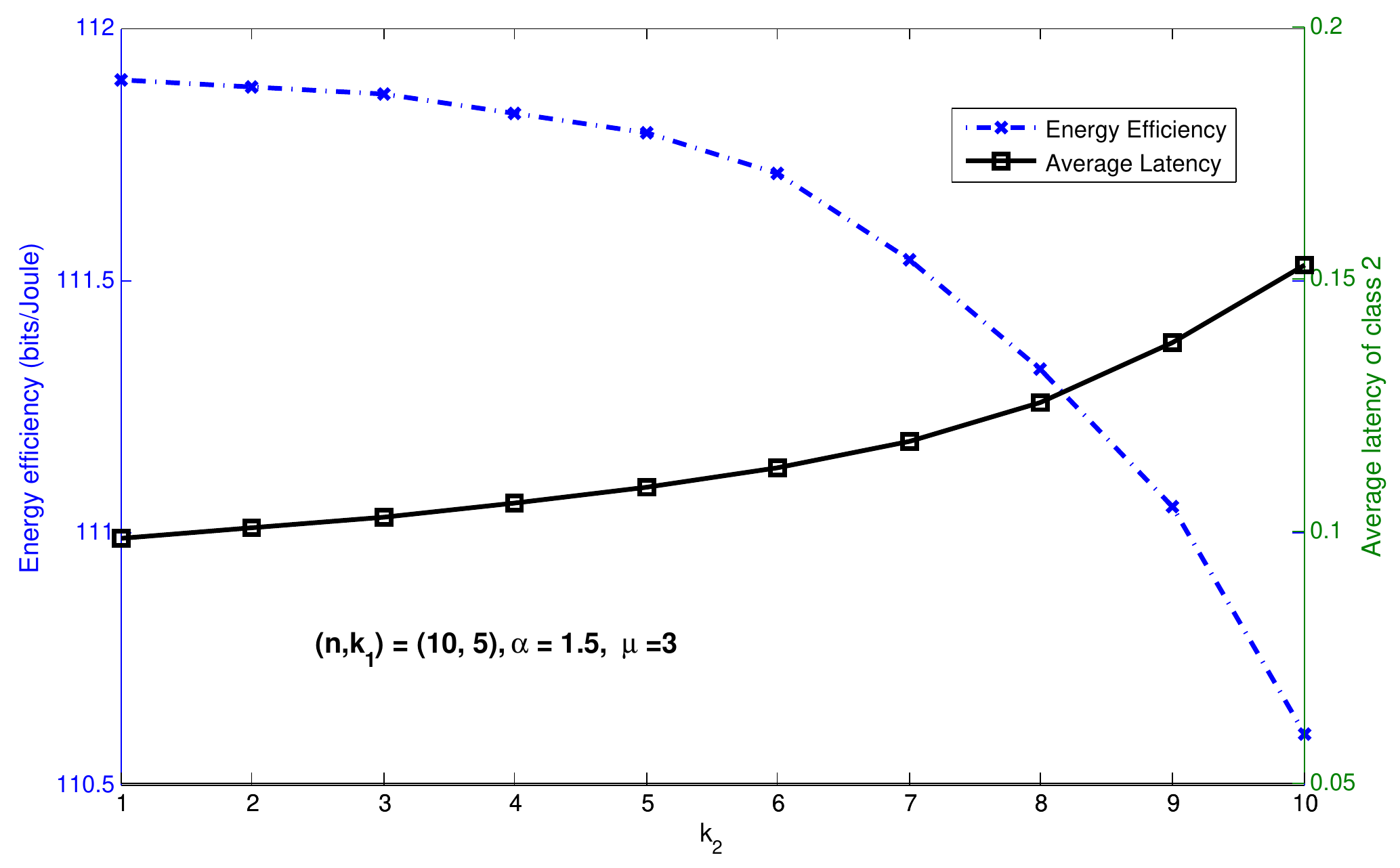}
\vspace{-5pt}
\caption{\footnotesize{A heavy tailed inter-arrival time distribution (Pareto-distribution with $\alpha=1.5$) results in monotonically increasing latency (and monotonically decreasing energy efficiency) as the code-rate is increased.}}
\vspace{-4pt}
\label{pareto_arrival}
\end{figure}

Fig.~\ref{pareto_arrival} illustrates the impact of a general (Pareto) arrival time distribution on the latency and energy efficiency of DSS. We observed that when distribution becomes heavy tailed, latency increases (and energy efficiency decreases) with increase in code rate. The heavy-tailed arrival distribution results in occasional very large inter-arrival time, however the arrival rate remains the same. Since it does not influence significantly the service dynamics, we observe that the latency increases with increase in code-rate similar to the M/M/1 case (in Fig.~\ref{muLow_class2}). Since latency increases, energy efficiency decreases with increase code-rate similar to previous results.

\subsection{Impact of number of redundant requests}\label{redReq}
\begin{figure}[!ht]
\centering
\includegraphics[scale=0.50]{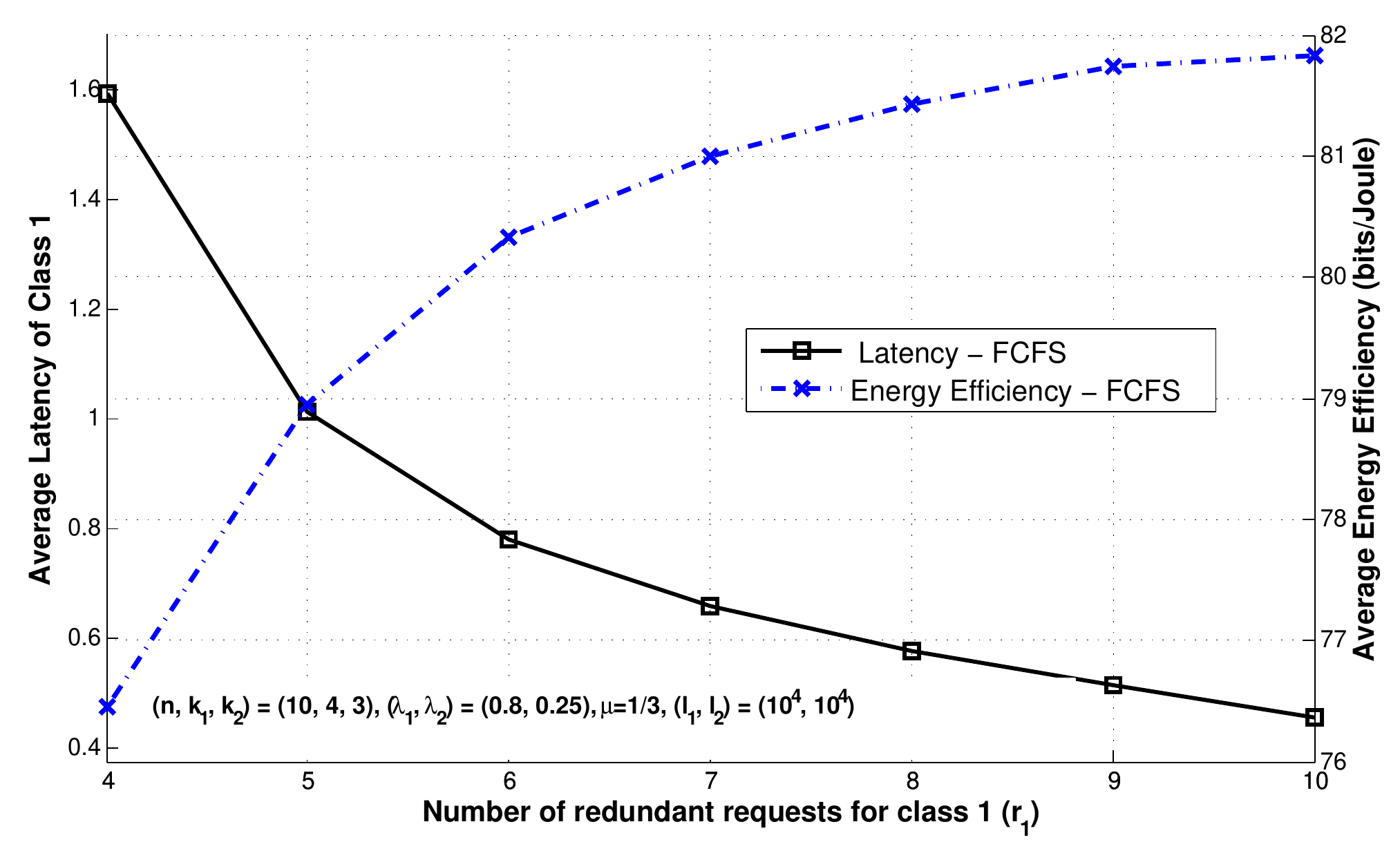}%
\vspace{-5pt}
\caption{\footnotesize{Sending redundant requests reduces average latency and improves energy efficiency.}}%
\label{varyR}
\end{figure}

We now explore the impact of varying the number of redundant requests (i.e., sending job requests to more than $k$ servers)  on the average latency and energy efficiency of DSS. The behavior of latency is governed by two opposing factors.
\begin{itemize}
\item Increasing the number of redundant requests reduces the service time because there are more servers available that simultaneously process the same job. This reduces the service time of each job. It increases the energy efficiency because the servers can process more requests per unit time. 
\item On the other hand, increasing the number of redundant requests reduces the number of servers available for serving the next job in queue, thus resulting in increase of size of queue at the servers. This results in loss of throughput and hence a plausible decrease in energy efficiency.
\end{itemize}
As it turns out that the first factor is more dominant than the second one, thereby resulting in an overall reduction in latency (increase in energy efficiency) by increasing the number of redundant requests. This behavior can be observed in
Fig.~\ref{varyR} which shows the average latency of class $1$ and energy efficiency of DSS for FCFS scheduling. In this figure, the redundancy for class $1$, $r_1$, is varied from $4$ to $10$ and the redundancy of class $2$ is set to $r_2 = 10$. 

\section{Conclusions}
In this paper, we proposed a novel multi-tenant DSS model and analyzed the energy efficiency of the system via lens of system latency. In the proposed heterogeneous DSS, each data class can possibly have different job arrival rate, job size and its data can be stored with a different fault-tolerance requirement by coding it with appropriate $(n,k)$ MDS code. In order to evaluate the impact of various parameters of DSS on its energy efficiency, we defined a \emph{data throughput} based energy efficiency metric for any given scheduling policy. We analytically established that the energy efficiency of DSS is inversely related to the system latency for a special case. This motivated us to further investigate the impact of various parameters on the relationship between the latency and energy efficiency of the DSS. Therefore, using a queuing-theoretic approach, we obtained bounds on the average latency for FCFS, preemptive and non-preemptive priority queuing policies. We verified the accuracy of the bounds for different settings of system parameters. The bounds, in general, are tight at high values of service rate, $\mu$ and low values of $k$. We also noted that the proposed lower bounds are tighter than a naive lower bound that follows directly from the work in \cite{gauri13}.

Using simulations, we investigate the relationship between average latency of data classes and energy efficiency of DSS under various setting of system parameters. We found that increasing the coding rate reduces the network bandwidth but increases latency and decreases energy efficiency. We also found that there exists an optimal number of servers which maximizes energy efficiency and results in near minimal latency. We observed that for heavy-tailed service distribution (which is the case for practical DSS), there exists an optimal code-rate that yield maximum energy efficiency and minimum latency. Lastly, we studied the impact of sending redundant requests on the average latency of that data class and the energy efficiency of DSS. We noted that increasing redundancy for a data class helps to reduce its average latency and as a consequence, the overall latency decreases and energy efficiency of DSS increases.

\appendices
\section{Stability Condition and Bounds on Average Latency}\label{append1}
\subsection{Proof of Lemma 1: Stability Condition-Priority queuing scheme}
Consider any node in the $(n, k_1, k_2, ..., k_R)$ Fork-Join system. Jobs of class $r$ enter the queue with rate $\lambda_r$. Each new job of class $r$ exits the system when $k_r$ sub-tasks of that job are completed. The remaining $n-k_r$ sub-tasks are then cleared from the system. Thus for each job of class $r$, $\frac{(n-k_r)}{n}$ fraction of the sub-tasks are deleted and hence the effective arrival rate of jobs of class $r$ at any node is $\lambda^r_{\textnormal{eff}} = \lambda_r\left(1-\frac{n-k_r}{n}\right) = \frac{k_r \lambda_r}{n}$. The stability condition for a priority queue is that the overall server utilization should be less than 1. If the condition is violated, the queues belonging to a priority level lower than some limit $k$ will grow without bound \cite[Section 3.5.3]{gallager}. Mathematically the stability condition is,
\begin{align}
\sum\limits_{r=1}^R \rho_r < 1,
\label{stab_NprQu1}
\end{align}
where $\rho_r = \frac{\lambda^r_{\textnormal{eff}}}{\mu_r}$ is the server utilization factor for class $r$. 
Substituting the expression for $\rho_r$ with $\mu_r = \frac{f k_r \mu}{l_r}$ in \eqref{stab_NprQu1} and rearranging terms, we get the stability condition of each queue as 
\begin{align}
\sum\limits_{r=1}^R{\lambda_r l_r} < nf\mu.
\end{align}
\subsection{Proof of Theorem 1}
\subsubsection{Upper Bound - Non-preemptive priority scheme}
Let $S_i$ be a random variable representing the service time for a job of class $i$. Then the average latency for a job of class $i$ in a non-preemptive priority scheduling system\footnote{The decreasing order of class priority is $1>2>\cdots>R$.} is given by \cite[Eq.(3.83)]{gallager},
\begin{align}
T^i_{\textnormal{N-PQ}} &= \textnormal{E}[S_i] + \frac{\sum\limits_{r=1}^R{\lambda_r \textnormal{E}[S_r^2]}}{2(1-\sum\limits_{r=1}^{i-1} \lambda_r \textnormal{E}[S_r])(1-\sum\limits_{r=1}^i \lambda_r \textnormal{E}[S_r])}, \label{npq_ub2}\\
&= \textnormal{E}[S_i] + \frac{\sum\limits_{r=1}^R{\lambda_r  ({(\textnormal{E}[S_i])}^2+\textnormal{V}[S_i])}}{2(1-\sum\limits_{r=1}^{i-1} \lambda_r \textnormal{E}[S_r])(1-\sum\limits_{r=1}^i \lambda_r \textnormal{E}[S_r])}. \label{npq_ub1}
\end{align}
To obtain an upper bound on the average latency, we degrade the FJ system in the following manner. For a job of class $i$, the servers that have finished processing a sub-task of that job are blocked and do not accept new jobs until $k_i$ sub-tasks of that job have been completed. Then the sub-tasks at remaining $n-k_i$ servers exit the system immediately. For jobs of class $i$, this performance-degraded system can be modeled as a M/G/1 system where the distribution of the service process, $S_i$, follows $k_i^{\textnormal{th}}$ ordered statistics as described in Section~\ref{sec:upperBound}. Now for for any class $i$, the service time at each of the $n$ servers is exponential with mean $1/\mu_i$. So the mean and variance of $S_i$ are given by \eqref{eq:meanVarSi}. Substituting \eqref{eq:meanVarSi} in \eqref{npq_ub1}, we get the following upper bound on average latency:
\begin{align}
T^i_{\textnormal{N-PQ}} \leq \overbrace{\underbrace{\frac{H_{n-k_i,n}^1}{\mu_i}}_\text{service time} + \underbrace{\frac{\sum\limits_{r=1}^R{\lambda_r[H_{n-k_r,n}^2  + {(H_{n-k_r,n}^1)}^2]/{\mu_r}^2}}{2\left(1-\mathcal{S}_{i-1}\right)\left(1-\mathcal{S}_i\right)}}_\text{waiting time}}^\text{Average latency of degraded system},
\end{align}
where $\mathcal{S}_i = \sum\limits_{r=1}^i \rho_r H_{n-k_r,n}^1$ and $\rho_r = \lambda_r/\mu_r$. 
%%%%%%%%%%%%%%%%%%%%%%%%%%%%%%%%%%%%%%%%%%%%%%%%%%%
\subsubsection{Upper Bound-Preemptive priority scheme}
Let $S_i$ be a random variable representing the service time for a job of class $i$. Then the average latency for a job of class (priority level) $i$ in a preemptive priority scheduling system\footnotemark[\value{footnote}] is given by \cite[Eq.(3.87)]{gallager},
\begin{align}
T^i_{\textnormal{PQ}} &= \frac{\textnormal{E}[S_i]}{1-\sum\limits_{r=1}^{i-1} \lambda_r \textnormal{E}[S_r]} + \frac{\sum\limits_{r=1}^i{\lambda_r \textnormal{E}[S_r^2]}}{2\left(1-\sum\limits_{r=1}^{i-1} \lambda_r \textnormal{E}[S_r]\right)\left(1-\sum\limits_{r=1}^i \lambda_r \textnormal{E}[S_r]\right)}, \label{pq_ub2} \\
&= \frac{\textnormal{E}[S_i]}{1-\sum\limits_{r=1}^{i-1} \lambda_r \textnormal{E}[S_r]} + \frac{\sum\limits_{r=1}^i \lambda_r \left({(\textnormal{E}[S_i])}^2+\textnormal{V}[S_i]\right)}{2\left(1-\sum\limits_{r=1}^{i-1} \lambda_r \textnormal{E}[S_r]\right)\left(1-\sum\limits_{r=1}^i \lambda_r \textnormal{E}[S_r]\right)}. \label{pq_ub1}
\end{align}

To obtain an upper bound on the average latency, we degrade the FJ system in the following manner. For a job of class $i$, the servers that have finished processing a sub-task of that job are blocked and do not accept new jobs until $k_i$ sub-tasks of that job have been completed. Then the sub-tasks at remaining $n-k_i$ servers exit the system immediately. For jobs of class $i$, this performance-degraded system can be modeled as a M/G/1 system where the distribution of the service process, $S_i$, follows $k_i^{\textnormal{th}}$ ordered statistics as described in Section~\ref{sec:upperBound}. Now for any class $i$, the service time at each of the $n$ servers is exponential with mean $1/\mu_i$. Hence the mean and variance of $S_i$ are given by \eqref{eq:meanVarSi}. Substituting \eqref{eq:meanVarSi} in \eqref{pq_ub1}, we get the following upper bound on average latency:
\begin{align}
T^i_{\textnormal{PQ}} \leq \overbrace{\underbrace{\frac{H_{n-k_i,n}^1}{\mu_i \left(1-\mathcal{S}_{i-1}\right)}}_\text{service time} + \underbrace{\frac{\sum\limits_{r=1}^i{\lambda_r[H_{n-k_r,n}^2  + {(H_{n-k_r,n}^1)}^2]/{\mu_r}^2}}{2\left(1-\mathcal{S}_{i-1}\right)\left(1-\mathcal{S}_i\right)}}_\text{waiting time}}^\text{Average latency of degraded system}.
\end{align}
where $\mathcal{S}_i = \sum\limits_{r=1}^i \rho_r H_{n-k_r,n}^1$ and $\rho_r = \lambda_r/\mu_r$. 

\subsection{Proof of Theorem 2}
\subsubsection{Lower Bound - Non-preemptive priority scheme}
For the purpose of obtaining a lower bound on the average latency of class $i$, using insights from Section~\ref{sec:lowerBound}, we map the parallel processing in the heterogeneous FJ system to a sequential process consisting of $k_i$ processing stages for $k_i$ sub-tasks of a job of class $i$. The transition from one stage to the next occurs when one of the remaining servers finishes a sub-task of the job.  Since classes are relabeled such that $k_1 \le k_2 \le ... \le k_R$, all jobs of classes 1 to $i$ get finished when stage $k_i$ is finished. However due to this relabeling of classes, this new class $1$ is not necessarily the one with highest priority.

\begin{figure}%
\centering
\includegraphics[scale = 0.40]{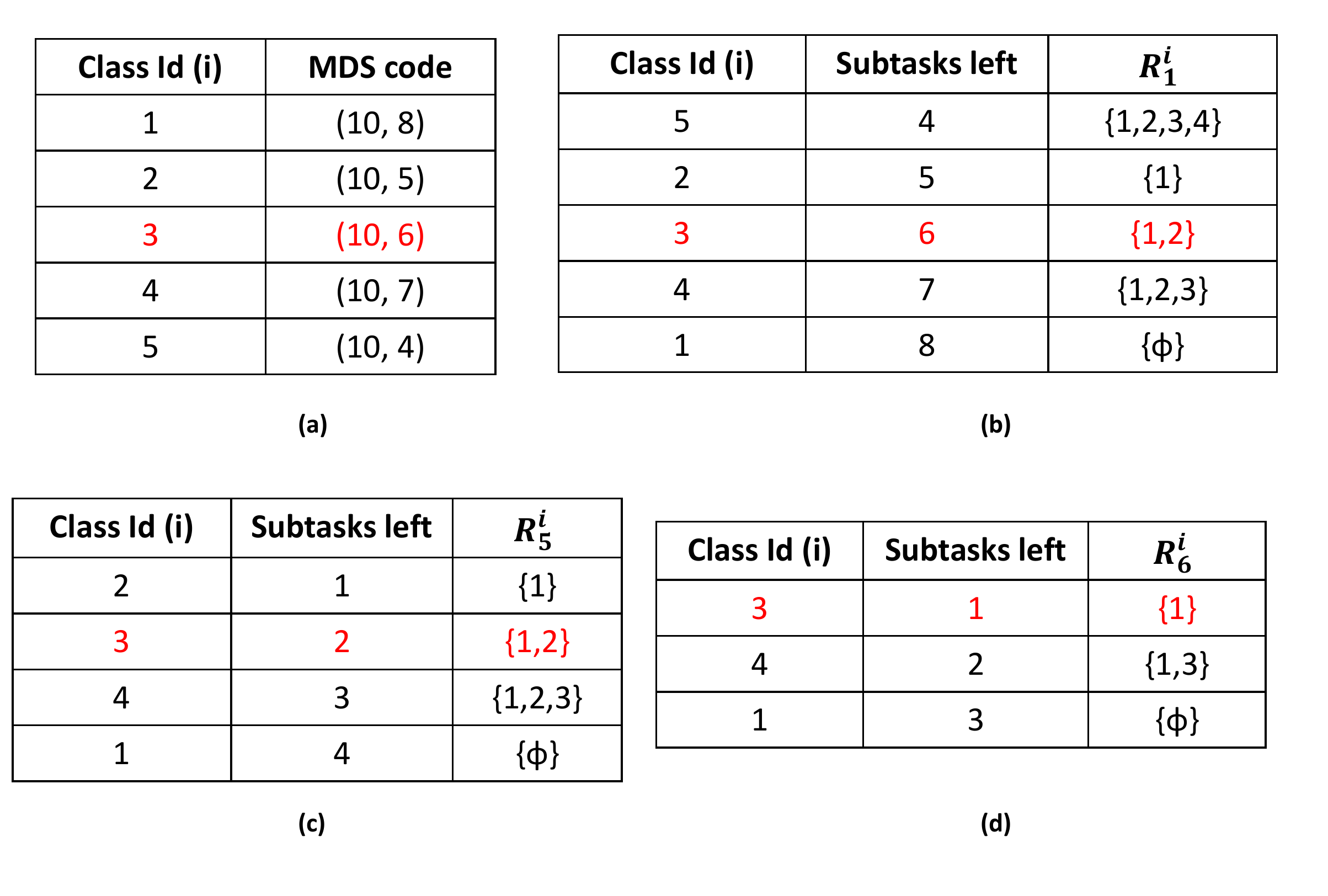}%
\vspace{-7pt}
\caption{Example illustrating the operational meaning of variable $\mathcal{R}_s^i$, where $i$ is the class id and $s$ is the processing stage. The decreasing order of class priority is $1>2>3>4>5$.}%
\label{exLBproof}%
\end{figure}

Let $c_s$ denote the number of classes that are finished before start of stage $s$, given by \eqref{csdef}. At any stage $s$, let $\mathcal{R}_s^{i}$ denote the set of classes with priority higher than class $i$ and have atleast one sub-task remaining to be completed. Fig.~\ref{exLBproof} illustrates the operational meaning of $\mathcal{R}_s^{i}$ using a toy example. Fig. 13(a) specifies the MDS codes corresponding to 5 data classes and suppose we are interested in the latency of class 3 marked in red. Fig. 13(b) shows the state of the system in stage $s=1$. The classes are reordered in order of increasing $k$ values. Since $s=1$, no class is finished yet and $c_1 = 0$. The last column shows the set $\mathcal{R}_1^i$ for a class $i$ with unfinished jobs. Fig. 13(c) shows the state of system at stage $s=5$. Since $k_5 = 4$, all jobs of class 5 are finished and exits the system, so $c_5 = 1$. However since class 5 has lowest priority, $\mathcal{R}_5^i$ for a remaining class $i$ is same as $\mathcal{R}_1^i$. Fig. 13(d) shows the state of the system at stage, $s=6$. Since $k_2 = 5$, all jobs of class 2 are finished and exits the system, so $c_6 = 2$. Since class 2 is higher priority than class 3, $\mathcal{R}_6^3$ is now $\{1\}$.

Now using \eqref{npq_ub2}, we get the mean completion time for sub-task of a job of class $i$ in stage $s$ as,
\begin{align}
T_{\textnormal{N-PQ},s,c_s,\mathcal{R}_s^i}^{i} &= \textnormal{E}[S_i^s] + \frac{\sum\limits_{r=c_s+1}^R{\lambda_r \textnormal{E}[{(S_r^s)}^2]}}{2\left(1-\sum\limits_{r \in \mathcal{R}_s^i} \lambda_r \textnormal{E}[S^s_r]\right)\left(1-\lambda_i \textnormal{E}[S_i^s]-\sum\limits_{r \in \mathcal{R}_s^i} \lambda_r \textnormal{E}[S_r^s]\right)}, \label{npq_lb}
\end{align}
where $S_i^s$ is a random variable denoting the service time for a sub-task of class $i$ in stage $s$. Unlike \eqref{npq_ub2}, the summation in the numerator starts from $c_s+1$ because classes 1 through $c_s$ have been completed at stage $s$. Also unlike \eqref{npq_ub2}, the summation in the denominator is over the set $\mathcal{R}_s^i \subseteq \{1, 2, \cdots, i-1\}$ because some of the higher priority (relative to class $i$) classes may have been finished at start of stage $s$.

Now at any stage $s$, the maximum possible service rate for a job of class $j$ that is not finished yet is $(n-s+1)\mu_j$. This happens when all the remaining sub-tasks of job of class $j$ are at the head of their buffers. Thus, we can enhance the latency performance in each stage $s$ by approximating it with a M/M/1 system with service rate $(n-s+1)\mu_j$ for jobs of class $j$. The average latency for sub-task of job of class $i$ in stage $s$ is thus lower bounded as,
\begin{align}
T_{\textnormal{N-PQ},s,c_s,\mathcal{R}_s^i}^{i} &\geq \overbrace{\underbrace{\frac{1}{(n-s+1)\mu_i}}_\text{service time} + \underbrace{\frac{\sum\limits_{r=c_s+1}^R{\frac{\lambda_r }{(n-s+1)\mu_r^2}}}{\left(1-\sum\limits_{r \in \mathcal{R}_s^i} \frac{\lambda_r}{(n-s+1)\mu_r}\right)\left(1-\frac{\lambda_i}{(n-s+1)\mu_i}-\sum\limits_{r \in \mathcal{R}_s^i} \frac{\lambda_r}{(n-s+1)\mu_r}\right)}}_\text{waiting time}}^\text{Average latency of enhanced system}.\nonumber
%\label{latEnhcdNPQ}
\end{align} 
 Finally, the average latency for class $i$ in this enhanced system is simply $\sum\limits_{s=1}^{k_i} T_{\textnormal{N-PQ},s,c_s,\mathcal{R}_s^i}^i$. Thus we have,
\vspace{-4pt}
\begin{align}
T^i_{\textnormal{N-PQ}} \geq  \sum\limits_{s=1}^{k_i} \left(\frac{t_{s,i}}{\lambda_i} + \frac{\sum\limits_{r=c_s+1}^R \frac{t_{s,r}}{(n-s+1)\mu_r}}{\mathcal{Z}_s^i\left(\mathcal{Z}_s^i-t_{s,i}\right)}\right),
\end{align}
%\vspace{-4pt}
where $t_{s,i} = \frac{\lambda_i}{(n-s+1)\mu_i}$ and $\mathcal{Z}_s^i = 1-\sum\limits_{r\in \mathcal{R}_{s}^{i}} t_{s,r}$.
%%%%%%%%%%%%%%%%%%%%%%%%%%%%%%%%%%%%%%%%%%%%%%%%%%%
\subsubsection{Lower Bound - Preemptive priority scheme}
For the purpose of obtaining a lower bound on the average latency of class $i$, using insights from \cite{varki}, we map the parallel processing in the heterogeneous FJ system to a sequential process consisting of $k_i$ processing stages for $k_i$ sub-tasks of a job of class $i$. The transition from one stage to the next occurs when one of the remaining servers finishes a sub-task of the job.  Since classes are relabeled such that $k_1 \le k_2 \le ... \le k_R$, all jobs of classes 1 to $i$ get finished when stage $k_i$ is finished. However due to this relabeling of classes, this new class $1$ is not necessarily the one with highest priority.

Let $c_s$ denote the number of classes that are finished before start of stage $s$, given by ~\eqref{csdef}. At any stage $s$, let $\mathcal{R}_s^i$ denote the set of classes with priority higher than class $i$ and have atleast one sub-task remaining to be completed. The operational meaning of $R_s^i$ is explained in Fig.~\ref{exLBproof}. Then using \eqref{pq_ub2}, the mean completion time for sub-task of a job of class $i$ in stage $s$ is given by,
\begin{align}
T_{\textnormal{PQ},s,c_s,\mathcal{R}_s^i}^{i} &= \frac{\textnormal{E}[S_i^s]}{1-\sum\limits_{r \in \mathcal{R}_s^i} \lambda_r \textnormal{E}[S^s_r]} + \frac{\lambda_i \textnormal{E}[{(S_i^s)}^2] + \sum\limits_{r \in \mathcal{R}_s^i} \lambda_r \textnormal{E}[{(S_r^s)}^2]}{2\left(1-\sum\limits_{r \in \mathcal{R}_s^i} \lambda_r \textnormal{E}[S^s_r]\right)\left(1-\lambda_i \textnormal{E}[S_i^s]-\sum\limits_{r \in \mathcal{R}_s^i} \lambda_r \textnormal{E}[S_r^s]\right)}, \label{pq_lb}
\end{align}
where $S_i^s$ is a random variable denoting the service time for a sub-task of class $i$ in stage $s$. Unlike \eqref{pq_ub2}, the summation terms in the numerator and denominator are over the set $\mathcal{R}_s^i \subseteq \{1, 2, \cdots, i-1\}$ because some of the higher priority (relative to class $i$) classes may have been finished at start of stage $s$. 

Now we note that at any stage $s$, the maximum possible service rate for a job of class $j$ that is not finished yet is $(n-s+1)\mu_j$. This happens when all the remaining sub-tasks of job of class $j$ are at the head of their buffers. Thus, we can enhance the latency performance in each stage $s$ by approximating it with a M/M/1 system with service rate $(n-s+1)\mu_j$ for jobs of class $j$. The average latency for sub-task of job of class $i$ in stage $s$ is thus lower bounded as,
\begin{align}
T_{\textnormal{PQ},s,c_s,\mathcal{R}_s^i}^{i} &\geq \overbrace{\underbrace{\frac{1}{(n-s+1)\mu_i\left(1-\sum\limits_{r \in \mathcal{R}_s^i} \frac{\lambda_r}{(n-s+1)\mu_r}\right)}}_\text{service time} + \underbrace{\frac{\frac{\lambda_i}{(n-s+1)\mu_i^2}+\sum\limits_{r \in \mathcal{R}_s^i}\frac{\lambda_r }{(n-s+1)\mu_r^2}}{\left(1-\sum\limits_{r \in \mathcal{R}_s^i} \frac{\lambda_r}{(n-s+1)\mu_r}\right)\left(1-\frac{\lambda_i}{(n-s+1)\mu_i}-\sum\limits_{r \in \mathcal{R}_s^i} \frac{\lambda_r}{(n-s+1)\mu_r}\right)}}_\text{waiting time}}^\text{Average latency of enhanced system}.\nonumber
%\label{latEnhcdPQ}
\end{align} 
\vspace{-4pt}
 Finally, the average latency for class $i$ in this enhanced system is simply $\sum\limits_{s=1}^{k_i} T_{\textnormal{PQ},s,c_s,\mathcal{R}_s^i}^i$. Thus we have,
\begin{align}
T^i_{\textnormal{PQ}} \geq  \sum\limits_{s=1}^{k_i} \left(\frac{t_{s,i}}{\lambda_i \mathcal{Z}_s^i} + \frac{1-\mathcal{Z}_s^i+\frac{t_{s,i}}{(n-s+1)\mu_i}}{\mathcal{Z}_s^i\left(\mathcal{Z}_s^i-t_{s,i}\right)}\right),
\end{align}
where $t_{s,i} = \frac{\lambda_i}{(n-s+1)\mu_i}$ and $\mathcal{Z}_s^i = 1-\sum\limits_{r\in \mathcal{R}_{s}^{i}} t_{s,r}$.
\vspace{-2pt}
\bibliographystyle{IEEETran}	
\bibliography{TCCarxiv}	

\end{document}